\documentclass[conference]{IEEEtran}
\IEEEoverridecommandlockouts
\usepackage{cite}
\usepackage{amsmath,amssymb,amsfonts}
\usepackage[utf8]{inputenc}
\usepackage{algorithmic}
\usepackage{graphicx}
\usepackage{textcomp}
\usepackage{xcolor}
\usepackage{color}
\usepackage{url}
\usepackage{bbm}
\usepackage{algorithm}
\usepackage{algorithmic}
\usepackage{bm}
 \usepackage{enumitem}
 \usepackage{amsthm}
 \usepackage{longtable}
 \usepackage[small]{caption}
 \usepackage{tabularx}
 \usepackage{multirow}

\newtheorem{thm}{Theorem}
\newtheorem{lem}{Lemma}
\newtheorem{cor}{Corollary}

\def\BibTeX{{\rm B\kern-.05em{\sc i\kern-.025em b}\kern-.08em
    T\kern-.1667em\lower.7ex\hbox{E}\kern-.125emX}}
\begin{document}

\title{Joint D2D Collaboration and Task Offloading for Edge Computing: A Mean Field Graph Approach 
}

\author{\IEEEauthorblockN{Xiong Wang\IEEEauthorrefmark{1},
Jiancheng Ye\IEEEauthorrefmark{2},
John C.S. Lui\IEEEauthorrefmark{1}}
\IEEEauthorblockA{\IEEEauthorrefmark{1} Department of Computer Science and Engineering, The Chinese University of Hong Kong}
\IEEEauthorblockA{\IEEEauthorrefmark{2} Network Technology Lab and Hong Kong Research Center, Huawei Technologies Co., Ltd.}
\IEEEauthorblockA{E-mail: xwang@cse.cuhk.edu.hk, yejiancheng@huawei.com, cslui@cse.cuhk.edu.hk}
}

\maketitle

\begin{abstract}
Mobile edge computing (MEC) facilitates computation offloading to edge server, as well as task processing via device-to-device (D2D) collaboration. Existing works mainly focus on centralized network-assisted offloading solutions, which are unscalable to scenarios involving collaboration among massive users. In this paper, we propose a \emph{joint framework} of decentralized D2D collaboration and efficient task offloading for a large-population MEC system. Specifically, we utilize the power of two choices for D2D collaboration, which enables users to beneficially assist each other in a decentralized manner. Due to short-range D2D communication and user movements, we formulate a \emph{mean field model} on a finite-degree and dynamic graph to analyze the state evolution of D2D collaboration. We derive the existence, uniqueness and convergence of the state stationary point so as to provide a \emph{tractable collaboration performance}. Complementing this D2D collaboration, we further build a \emph{Stackelberg game} to model users’ task offloading, where edge server is the leader to determine a service price, while users are followers to make offloading decisions. By embedding the Stackelberg game into Lyapunov optimization, we develop an \emph{online} offloading and pricing scheme, which could optimize server's service utility and users' system cost simultaneously. Extensive evaluations show that our  D2D collaboration can  mitigate users' workloads by $73.8\%$ and  task offloading can achieve high energy efficiency.
\end{abstract}
\section{Introduction}
In recent years, we have witnessed  a rapid growth of data generated from the network edge, especially with the enormous popularity of mobile devices~\cite{the2016satyanarayanan}. Many intelligent mobile applications, such as interactive gaming,  real-time face recognition and natural language processing,  are emerging which typically demand intensive computation and low latency. In general, mobile devices have constrained resources, while remote-resided cloud server suffers from high latency due to long-haul transmissions. To support the compute-intensive yet delay-sensitive applications, mobile edge computing (MEC) is recognized as a new paradigm to push cloud frontier close to the edge for such  service requirements~\cite{edge2016shi}. 

%
At a high level, MEC enables mobile users to offload tasks to the local edge server endowed with computing functionalities. Compared to the cloud datacenter, an individual edge server basically has limited computing capacity, making it difficult to accommodate huge amount of tasks since  over 90\% of the data will be stored and processed at the network edge~\cite{kelly2020internet}. Under this scenario, exploiting \emph{collaboration among users} is a promising complementary approach to task offloading for  easing the strain on the edge server~\cite{chen2017exploiting}. Device-to-device (D2D) communication (e.g., via Bluetooth or Wi-Fi Direct) generally is more energy-saving while less time-consuming~\cite{asadi2014a}, thereby providing a low-latency service  for users when they collaboratively process tasks via D2D links~\cite{pu2016D2D}. Specifically,  heavily-loaded users can seek immediate assistance from  lightly-loaded ones within proximity, and hence the average task delay is expected to decrease significantly. 

Despite the clear advantage of D2D collaboration, task offloading to edge server is still indispensable for MEC as high latency is witnessed if compute-intensive tasks are handled solely by  resource-constrained mobile devices. Along with the offloading, a service price is charged by the edge server when providing  computing service for mobile users~\cite{zhao2020intelligent}. Needless to say, setting a proper price is critical, as an excessively low price is insufficient to compensate for the server's operation cost, whereas an unduly high price will certainly cause a decrease in user demands of task offloading and further increase the delay in task execution. Therefore, a reasonable pricing scheme is required to incentivize task offloading while also bringing benefits to the edge server operator. 

Various efforts have been dedicated to investigate D2D collaboration in MEC or fog computing~\cite{chen2017exploiting,pu2016D2D,xing2019joint}. Benefiting from  users' mutual assistance, D2D collaboration can effectively improve energy efficiency and  delay performance for MEC system. In these works, the edge server mainly serves as a central coordinator to aid the collaborative task processing, whereas the potential benefit of task offloading was not explored. Few following works further incorporate D2D collaboration into task offloading to take advantage of the computing capacity embedded in mobile devices and edge server~\cite{chen2018socially,he2019D2D}. However, these works mostly concentrate on a centralized offloading and/or collaboration optimization, with a restrictive assumption of   time-invariant D2D links in order to achieve a tractable analysis.  When there are  a large population of moving mobile users, which is often the case in D2D collaboration, \emph{how to characterize a decentralized collaboration  and develop an efficient offloading for ``dynamic'' MEC system} still remains unresolved. To answer this critical question, researchers are faced with the following challenges.

First, due to short-range D2D communication, collaboration occurs mainly among nearby users, thus leading to a graph structure formed by spatially distributed mobile users. A decentralized collaboration scheme should encompass both \emph{static and dynamic connectivity setting} when considering user movements, which however is theoretically challenging in general. Therefore, it brings new modeling requirements for D2D collaboration to achieve rigorous theoretical guarantee as well as  good empirical performance. Second, task offloading is influenced by the service price set by edge server, whereas the pricing scheme is also dependent on the strategic offloading decision of mobile users. Their \emph{mutual-dependency} raises difficulty in the optimal offloading and pricing design. Third, D2D collaboration is \emph{intertwined} with task offloading due to various task executions including collaborative execution and offloaded execution. This demands incorporating decentralized collaboration into determining appropriate proportions of tasks to be offloaded and to be processed locally so as to reduce the execution latency while enhancing energy efficiency.

In this paper, we propose a \emph{joint D2D collaboration and task offloading} for a large-population MEC system. We first use the power of two (Po2) choices to enable a \emph{decentralized collaboration} among massive mobile users, where each user randomly polls a neighbor within its D2D range and forwards a task if the polled neighbor has a lighter workload.  We develop a novel \emph{mean field  model on graph}  to analyze this D2D collaboration, through which we can characterize the state evolution of MEC system in both static and dynamic situations. By incorporating the steady state of D2D collaboration, we further formulate a Stackelberg game to model the task offloading from mobile users to edge server. Specifically, users are followers in making their offloading decisions, while  edge  server is the leader in determining a dynamic price based on \emph{Lyapunov optimization}  for providing computing service. As a result, we consider the intertwined collaboration and offloading processes so as to collectively promote an efficient task execution. This paper has the following main contributions:
\begin{itemize}[leftmargin=*]
\item We develop a joint D2D collaboration and  task offloading framework which facilitates users to collaboratively process tasks in a \emph{decentralized manner} and offload computation to local edge server. Our framework targets the real-life large-population MEC system so as to fully unleash the potentials of widely-distributed mobile devices and the edge server's capacity. Evaluations show that we can reduce users' workloads by $73.8\%$ and improve the energy efficiency as well. 
\item We propose a novel mean field model on static and dynamic graphs to characterize D2D collaboration, based on which we can analyze the stochastic state evolution by \emph{deterministic ordinary differential equations} (ODEs). We rigorously prove the existence and uniqueness of mean field stationary point to provide a \emph{theoretically tractable performance} for D2D collaboration. To the best of our knowledge, this is the first work that conducts a thorough analysis of mean field model on \emph{finite-degree and dynamic graphs}.
\item We design an \emph{online offloading and pricing scheme} using a Lyapunov optimization framework to determine the optimal offloading and pricing decisions over time. By embedding a Stackelberg game into the online decision making,  we can simultaneously minimize users' system cost while meeting their stringent task delay requirements, and maximize the server's \emph{long-term utility} with only current information.
\end{itemize}

\begin{figure}[t]
  \centering
  \includegraphics[width=0.88\columnwidth, height = 1.115in]{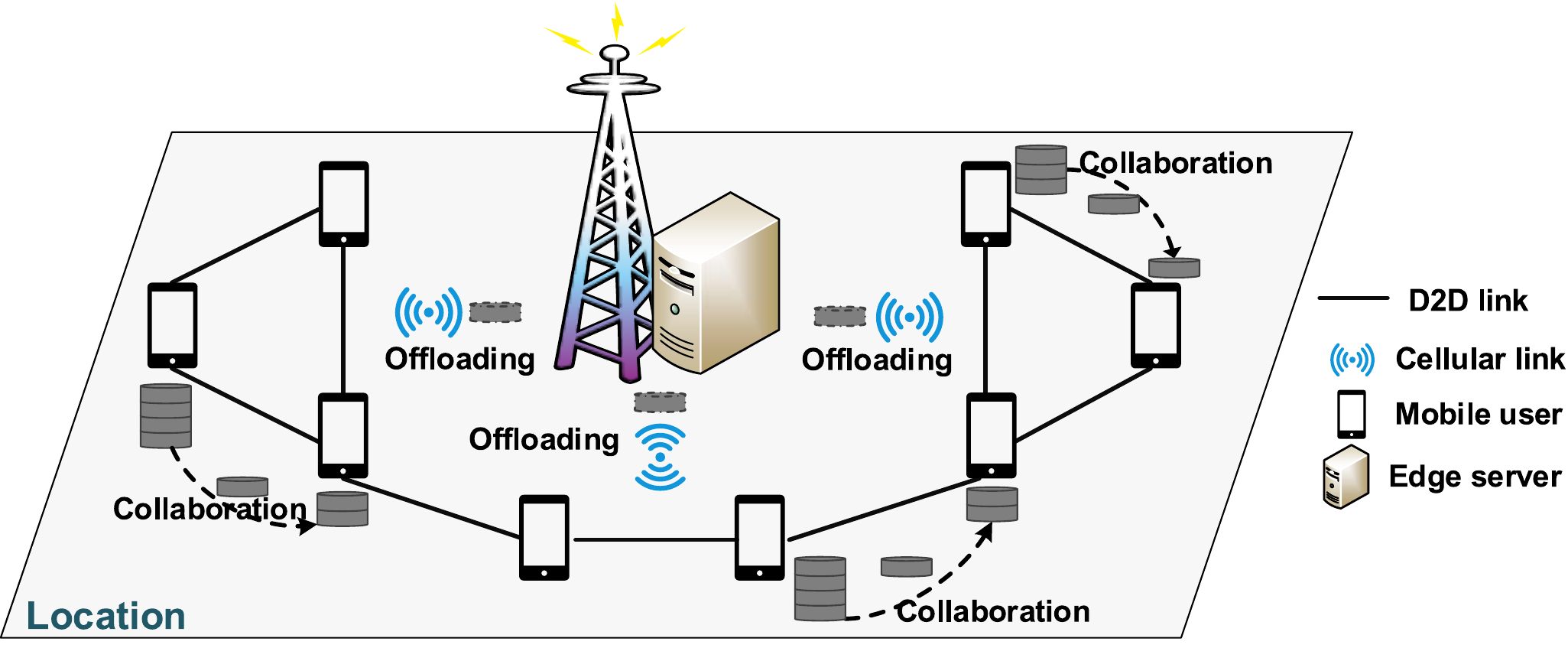}
  \caption{Snapshot of the MEC system.}\label{Fig:sysmodel}
  \vspace{-15pt}
\end{figure}

\section{System Model} \label{Sec:Sys_Model}
We consider a MEC system with  large-population mobile users  $\mathcal{N}=\{1,2,...,N\}$ and an edge server. Users' tasks can be offloaded to edge server via cellular network or processed by their collaboration via D2D link, as shown in Fig.~\ref{Fig:sysmodel}. 
\subsection{System Overview}
Due to short-range D2D communication, collaboration is mainly among users within proximity,  and we model this collaboration structure as a connected graph $\mathcal{G} = \{\mathcal{N}, \mathcal{E}\}$ with $\mathcal{E}$ denoting the D2D links.  Besides, along with task offloading, a price $p$ is charged by the edge server for providing computing service, where $p$ \emph{remains fixed} for a long period, e.g., weekly or monthly basis~\cite{zhang2014time}. In return, users strategically choose to offload tasks with probability $x \in [0,1]$, while handling the rest via D2D collaboration. A task is typically characterized by the required service time (CPU cycles) and the amount of cellular traffic (data size)~\cite{kwak2015dream}.  By convention, task generation of each user follows a rate-$\lambda$ Poisson process, where the service time of a task obeys an exponential distribution with normalized unit mean value, and the data size has an average value of $B$~\cite{chen2018peer}. The normalized service rates of mobile devices and edge server are $\mu$ and $\gamma$, respectively, with $\lambda < \mu$ to keep the MEC system stable. Moreover, the server can process the offloaded tasks in parallel because it has a more powerful computing capability than mobile devices~\cite{cardellini2016a}. 

\subsection{D2D Collaboration}
We use the Po2 choices for decentralized collaboration~\cite{mitzenmacher2001the}. Let $Q_u(t)$ be the number of tasks, or workload, of user $u \in \mathcal{N}$ at time $t$. For Po2, when a task is generated by $u$ and not offloaded, $u$ randomly polls a neighbor, say $v$, and forwards the task to $v$ if $Q_u(t) > Q_v(t)$; otherwise the task joins $Q_u(t)$ with ties being broken arbitrarily. Denote $d_u$ as the number of $u$'s neighbors, also known as its degree in graph  $\mathcal{G}$. The degree $d_u$ is distributed in a \emph{finite degree set} $\mathcal{K} = \{k_{\min},...,k_{\max}\}$ due to short-range D2D communication.  W.l.o.g., the graph  $\mathcal{G}$ is  \emph{uncorrelated}~\cite{pastor2015epidemic}, i.e., the probability  $p(k'|k)$ that a user with degree $k$ has a link to a neighbor with degree $k'$ satisfies:
\begin{equation} \label{Eq:Uncor}
p(k'|k) = \frac{k'p(k')}{\overline{k}},
\end{equation}
where $p(k)$ is the probability that  a user has degree $k$ and $\overline{k} = \mathbb{E}[k]=\sum_{k \in \mathcal{K}} p(k)k$ is the expected degree. Eq.~\eqref{Eq:Uncor} means the degree distribution of two neighboring users is  independent. 

For $k \in \mathcal{K}, i \ge 0$, we define $q_{k,i}(t)$  and $s_{k,i}(t)$ as:
\begin{equation} \label{Eq:Por-q}
q_{k,i}(t) =\frac{ \sum_{u \in \mathcal{N}} \mathbf{1}_{d_u=k, Q_u(t)=i}}{\sum_{u \in \mathcal{N}} \mathbf{1}_{d_u=k}},
\end{equation}
\begin{equation}\label{Eq:Por-s}
s_{k,i}(t) = \sum \nolimits_{j \ge i} q_{k,j}(t).
\end{equation}
Physically, $q_{k,i}(t)$ or $s_{k,i}(t)$ can be regarded as the probability that a  user with degree $k$ holds $i$ or at least $i$ tasks, respectively. Also, $q_{k,i}(t) = s_{k,i}(t) - s_{k,i+1}(t)$ and $s_{k,0}(t) = 1, \forall k \in \mathcal{K}$. We say  $s_{k,1}(t)$ is the \emph{busy probability} as it implies the case where a $k$-degree user has a non-empty workload. Denote $\bm{s}(t) = \{s_{k,i}(t)\}$ as  the system state, that is, the workload distribution of the MEC system. Hereinafter, we will use  $q_{k,i}, s_{k,i},Q_u$ and $\bm{s}$ without index $t$ if there is no confusion. 

Our objective is to demonstrate that D2D collaboration can effectively mitigate users' workloads, by characterizing the stationary point $\bm{s}^*$ for state evolution when user number $N \rightarrow \infty$ and  $\dot{\bm{s}}^* = \bm{0}$ (MEC system is stable). Given the graph collaboration structure, we develop a new analysis framework as existing Po2 methods are no longer applicable.



\subsection{Task Offloading} \label{Sec:Opt_price}
Upon generating a task, users offload it to the edge server with probability $x$, and are charged with a price $p$. Since $p$ is fixed for a long time, the offloading and pricing decisions are made  in terms of discrete time slot $\{0,1,...,n,...\}$, say at a time interval of every week. The probability $x[n]$ and price $p[n]$ are constant in each slot, and the MEC system is regarded to be stable. Hence, we can leverage the \emph{stationary point} $\bm{s}^*$ in offloading and pricing scheme design.

\subsubsection{Offloading constraint}
Together with D2D collaboration, users also collaboratively decide the probability $x[n]$ in each time slot, which actually only relies on the information of $\bm{s}^*$.

\textbf{Task delay}. If a task is offloaded to the server, the expected task delay includes  the transmission time and server processing time, that is $d_o= x[n](\frac{B}{r}+\frac{1}{\gamma})$ where $r$ denotes the data rate of cellular network and $1$ is the normalized unit service time. With probability $x_c[n] = 1- x[n]$, a task will be processed via D2D collaboration, then the delay amounts to the queueing time, which is $d_q=x_c[n]\frac{\sum_{i} \sum_{k}p(k)s_{k,i}^*}{x_c[n]\lambda} = \frac{\sum_{i}\sum_{k}p(k)s_{k,i}^*}{\lambda}$ based on  Little's law. Here, the transmission delay of  fast and short-range D2D communication is negligibly small compared to $d_o$ and $d_q$~\cite{asadi2014a}.  Therefore, the average task delay $d(x[n])$ is:
\begin{equation}\label{Eq:delay}
d(x[n])= d_o+d_q.
\end{equation}
Note that $s_{k,i}^*$ depends on $x_c[n]$, or offloading probability $x[n]$. 

\textbf{Collaboration fairness}. Due to heterogeneous number of neighbors, users have \emph{unbalanced contributions} in D2D collaboration, i.e., busy probability  $s_{k,1}^*$ varies over degree $k$. When deciding the probability  $x[n]$, collaboration fairness requires that the gap between the highest  and lowest $s_{k,1}^*$ should not be too large so as to prevent  the ``free-riding'' scenario.

\subsubsection{Pricing constraint} The offloading probability $x[n]$ of mobile users is  affected by the service price $p[n]$ charged by the edge server. In general, setting a high price will restrain user demands of task offloading, or low $x[n]$, whereas positing a low price will lead to \emph{overloaded situation} at the edge server because of too many offloaded tasks. We consider that  the server can  adaptively choose $p[n]$ for compensating  its operation cost and avoiding being overloaded.


\subsubsection{Problem formulation} We now formulate the system cost of users and the service utility of the server in task offloading.

\textbf{Users' system cost}. Since D2D communication is energy efficient, the system cost of a user is  mainly composed of charged fee, processing cost, and offloading transmission cost. Formally, the charged fee is the payment to the edge server for task offloading, which is $x[n] \lambda p[n]$. When processing a task, it needs an average $\frac{1}{\mu}$ time, so the energy consumption is $\rho_c^m \frac{1}{\mu}$ where $\rho_c^m$ is the energy cost per CPU cycle for computation in a mobile device~\cite{kwak2015dream}. Moreover, the expected busy probability is $s^*_1 \triangleq \sum_{k \in \mathcal{K}} p(k) s_{k,1}^*$ from Eq.~\eqref{Eq:Por-s}.  Therefore, the processing cost becomes $s^*_1 \frac{\rho_c^m}{\mu}$. Finally, the transmission cost is $x[n] \lambda \rho_t^m \frac{B}{r}$ where $\rho_t^m$ is the unit cost for transmitting cellular traffic. Overall, the system cost $c[n]$ in time slot $n$ is:
\begin{equation} \label{Eq:cost}
\vspace{-1pt}
c[n] =x[n] \lambda p[n] + s^*_1 \frac{\rho_c^m}{\mu} + x[n] \lambda \rho_t^m \frac{B}{r}.
\vspace{-2pt}
\end{equation}

\textbf{Server's service utility}. On the server's side, its average profit in time slot $n$ is $x[n] \lambda p[n]$, and average processing cost is $ x[n]  \lambda \frac{\rho_c^s}{\gamma}$ with $\rho_c^s$ representing the energy cost per CPU cycle in the edge server. Then, the service utility $u[n]$ is acquired:
\begin{equation} \label{Eq:utility}
u[n] =x[n] \lambda p[n] - x[n]  \lambda \frac{\rho_c^s}{\gamma}.
\vspace{-3pt}
\end{equation}

\textbf{Stackelberg game}. Given the price $p[n]$, users aim to reduce their system cost by deciding the probability $x[n]$, subject to constraints of task delay and collaboration fairness:
\begin{subequations} \label{Eq:cost_min}
\begin{align}
\min \nolimits_{x[n]}~~& c[n] \tag{\ref{Eq:cost_min}}\\
\mathrm{s.t.}~~& d(x[n]) \le \overline{d} \label{Eq:delayCon}\\
&\max \{s_{k,1}^*\}  - \min \{s_{k,1}^*\}  \le \overline{s} \label{Eq:fairCon}.
\vspace{-2pt}
\end{align}
\end{subequations}
As for the edge server, its objective is to optimize the long-term utility via dynamically setting the service price $p[n]$:
\begin{subequations} \label{Eq:utility_max}
\begin{align}
\max \nolimits_{p[n]}~~& \lim_{T \rightarrow \infty} \frac{1}{T} \sum \nolimits_{n=0}^{T-1} \mathbb{E}[u[n]]  \tag{\ref{Eq:utility_max}} \\
\mathrm{s.t.}~~& \lim_{T \rightarrow \infty} \frac{1}{T} \sum \nolimits_{n=0}^{T-1} \mathbb{E}[x[n] \lambda] \le \overline{x} \label{Eq:overload}\\
& p[n] \in (0,p_u] \label{Eq:accept},
\vspace{-2pt}
\end{align}
\end{subequations}
where the inequality in Eq.~\eqref{Eq:overload} is  the overloaded constraint and $p_u$ denotes the highest price users could accept. 

In time slot $n$, the server first chooses a price $p[n]$, and then users react via the offloading decision $x[n]$, which is modeled as a Stackelberg game. Note that optimizing cost and utility is intertwined with D2D collaboration. Also, we have to design an offloading and pricing scheme to simultaneously maximize service utility online and minimize system cost offline.

\section{Mean Field D2D Collaboration} \label{Sec:MFDynamics}
In this section, we formulate a mean field model on graph to analyze D2D collaboration. Specifically, we will derive the state evolution by allowing the number of users $N$ to approach infinity. Since users may move around, we  consider the decentralized collaboration  on both static and dynamic graphs to encompass  the case of \emph{time-varying D2D links}.  

\subsection{Collaboration on Static Graph}
Basically, a static graph implies that D2D links are time-invariant, i.e., the graph $\mathcal{G}$ remains unchanged throughout the collaboration. To characterize state $\bm{s}$, we explore  the transition of each $s_{k,i}$ from the perspective of a particular $k$-degree user $u$, as users are asymptotically independent when $N\rightarrow \infty$.

\subsubsection{State evolution}
Considering $s_{k,i}$ represents the workload distribution of $k$-degree users,  then  the  transition events for the Markov chain include  the following three instances:
\begin{itemize}[leftmargin=*]
\item The number of tasks is $Q_u = i-1$, and $u$ generates a task which stays at $u$, so the state transits from $s_{k,i-1}$ to $s_{k,i}$.
\item The number of tasks is $Q_u = i-1$, and  $u$  receives a task sent from a neighbor, then $s_{k,i-1}$ transits to $s_{k,i}$.
\item The number of tasks is  $Q_u =i$, and  $u$ processes a task locally. Hence,  the state changes from $s_{k,i}$ to $s_{k,i-1}$. 
\end{itemize}
Transition probability of each instance is now provided. For the first instance, the polled neighbor by $u$ must have no fewer tasks, which occurs with probability $q_{k,i-1}\sum_{k' \in \mathcal{K}}p(k'|k)s_{k',i} + q_{k,i-1}\frac{1}{2}\sum_{k' \in \mathcal{K}}p(k'|k)q_{k',i-1}$. The first term $q_{k,i-1}\sum_{k' \in \mathcal{K}}p(k'|k)s_{k',i}$ means $Q_u=i-1$ and the polled neighbor has at least $i$ tasks. The second term $q_{k,i-1}\frac{1}{2}\sum_{k' \in \mathcal{K}}p(k'|k)q_{k',i-1}$ denotes tie breaking, namely the polled user also holds $i-1$ tasks. As for the second instance that $u$ receives a task from a neighbor, its probability is $kq_{k,i-1} \sum_{k' \in \mathcal{K}}p(k'|k)s_{k',i} \frac{1}{k'} + kq_{k,i-1} \frac{1}{2}\sum_{k' \in \mathcal{K}}p(k'|k)q_{k',i-1} \frac{1}{k'}$, where $kq_{k,i-1}$ is because $u$ has $k$ neighbors and $ \frac{1}{k'}$ implies that a $k'$-degree neighbor randomly polls $u$ in Po2 choices (D2D collaboration)  with probability  $ \frac{1}{k'}$. Similarly, the two terms represent situations where $u$ has fewer tasks and tie breaks, respectively. The probability of the last instance is simply $q_{k,i}$. Combining task generation rate $\lambda$, offloading probability $x$ and service rate of a mobile device $\mu$, the state evolution is specified:
\begin{equation}  \label{Eq:state_dya}
\begin{aligned}
&\dot{s}_{k,i}=-\mu q_{k,i}+ x_c\lambda q_{k,i-1} \sum_{k' \in \mathcal{K}}p(k'|k) \Bigl(s_{k',i}+\frac{1}{2}q_{k',i-1}\Bigr)\\
&+ k x_c\lambda q_{k,i-1} \sum_{k' \in \mathcal{K}} \frac{1}{k'} p(k'|k) \Bigl(s_{k',i}+\frac{1}{2}q_{k',i-1}\Bigr),
\end{aligned}
\end{equation}
where $x_c = 1-x$ is the probability that a task is processed via D2D collaboration and $x$ would be $x[n]$ if in time slot $n$. 

\subsubsection{ODE system} Remember that  $q_{k,i}=s_{k,i}-s_{k,i+1}$ and $\mathcal{G}$ is an uncorrelated graph. Based on Eq.~\eqref{Eq:Uncor}, we have $p(k'|k) = \frac{k'p(k')}{\overline{k}}$. Therefore, for $i>0$,  Eq.~\eqref{Eq:state_dya} is simplified to:
\begin{equation} \label{Eq:ODE-s}
\begin{aligned}
&\dot{s}_{k,i}=-\mu(s_{k,i}-s_{k,i+1}) \\
&+x_c \lambda (s_{k,i-1}-s_{k,i})\Bigl[\frac{1}{2}\sum_{k'\in\mathcal{K}}\frac{k'+k}{\overline{k}}p(k')(s_{k',i-1}+s_{k',i})\Bigr].
\end{aligned}
\end{equation}
Besides, $s_{k,0}=1$ according to Eq.~\eqref{Eq:Por-s}. Define the drift function $\bm{F}(\bm{s}) = \left\{F_{k,i}(\bm{s}) \right\}$, where $F_{k,0}(\bm{s}) = 0$ and $F_{k,i}(\bm{s})=\dot{s}_{k,i}, \forall i>0$. We have the following form:
\begin{equation} \label{Eq:drift}
\dot{\bm{s}} = \bm{F}(\bm{s}).
\end{equation}
The \emph{deterministic ODE system} of Eq.~\eqref{Eq:ODE-s}  corresponds to the \emph{mean field model for our characterized D2D collaboration}. 



\subsection{Collaboration on Dynamic Graph}
As users may move around,  their neighbors within D2D communication range also change accordingly, resulting in a  time-varying graph structure. Specifically, we leverage the model in~\cite{casteigts2012time, lang2018analytic} to capture this dynamic feature. 

\subsubsection{Dynamic graph model} In a dynamic graph, each user has an expected degree $k$ which is fixed and drawn from a finite set  $\mathcal{K}=\{k_{\min},...,k_{\max}\}$ with probability $p(k)$. Value of expected degree indicates  the willingness of a user to participate in collaboration. Hence, a D2D link between two users is established based on their expected degrees and spatial distance. In this regard, the number of D2D neighbors, or realized degree, of a user follows certain distribution conditioned on its expected degree $k$, which is specified as a Poisson distribution with the mean value being $k$ in~\cite{ lang2018analytic}. Due to user mobility, graph  $\mathcal{G}(t)$ is dynamic with a time-varying edge set $\mathcal{E}(t)$, or changing realized degrees.  Similar to the static case, $\mathcal{G}(t)$ formed by realized degrees is considered uncorrelated.

\subsubsection{State evolution} Define $q_{k,i}, s_{k,i}$ as  Eqs.~\eqref{Eq:Por-q}-\eqref{Eq:Por-s}, whereas $k$ now denotes the \emph{expected degree}. Similar to the static graph, transitions of state $s_{k,i}$ on dynamic graph also entail  three instances. The main difference lies in the probability $p(k'|k)$, a user has expected degree $k$ and its neighbor has expected degree $k'$, is no longer the expression in Eq.~\eqref{Eq:Uncor}. Instead, we use the conditional distribution of the realized degree  to help compute this $p(k'|k)$.  Due to space limit, we elucidate the details of deriving the state evolution in Appendix~\ref{App:dynamicG}, and only present the final result here. We find out that the  evolution of $s_{k,i}$  is exactly  Eq.~\eqref{Eq:ODE-s}, i.e., \emph{the mean field models on static and dynamic graphs are unified by the same ODE system.} Hereinafter, we will rely on Eq.~\eqref{Eq:ODE-s} to analyze D2D collaboration in both static and dynamic scenarios as a whole. 



\subsection{Mean Field Model on Finite-degree Graphs}
There have been many efforts devoted to mean field model on graphs, while existing works mostly focus on complete and infinite-degree graphs~\cite{mitzenmacher2001the,budhiraja2017super}, or apply the mean field analysis without theoretical guarantees~\cite{pastor2015epidemic,gast2015the}. Therefore, our work has  two novel contributions in the mean field aspect. First, we extend current mean field model on graphs to both \emph{static and dynamic graphs with finite yet heterogeneous degrees}. Second, we also provide \emph{rigorous proofs} (in the next section), along with extensive evaluations to demonstrate the effectiveness of our mean field model. 

Now we discuss a special case where users have a homogeneous degree, that is the degree set $\mathcal{K}=\{k\}$ and $p(k)=1$. Consider that the graph $\mathcal{G}$ is connected and uncorrelated. The mean field model becomes (irrelevant to degree $k$ indeed):
\begin{equation} \label{Eq:ODE-Spe}
\dot{s}_{k,i}= x_c\lambda (s^2_{k,i-1}-s^2_{k,i})-\mu(s_{k,i}-s_{k,i+1}).
\end{equation}
This in fact degenerates to the classical Po2 result~\cite{mitzenmacher2001the}.  

Our main objective is to obtain the stationary point $\bm{s}^*$ for state evolution, such that $\bm{F}(\bm{s}^*) = \bm{0}$. To this end, we will derive the existence and uniqueness of the stationary point, as well as demonstrating that the system state from \emph{any initial point} will eventually converge to this stationary point. 

%


\section{Stationary Point for State Evolution} \label{Sec:MFIP}
In this section, we  specify the stationary point for the mean field model to obtain the steady state of D2D collaboration.


\subsection{Stationary Point}
Since drift at the stationary point is $0$, the MEC system is  statistically stable, and hence we are able to  achieve a tractable collaboration performance. For this reason, the existence and uniqueness issues of stationary point need to be explored.
\subsubsection{Existence of stationary point}
We first demonstrate that there exists a stationary point $\bm{s}^*$ for our mean field model. Considering $\bm{F}(\bm{s}^*) = \bm{0}$, we attain the existence by showing that the ODE system of Eq.~\eqref{Eq:ODE-s} has a fixed point.

\begin{thm} \label{Thm:exist}
There exists a stationary point $\bm{s}^*$ for the mean field model.
\end{thm}


See Appendix~\ref{App:exist} for the detailed proof. With the existence of  stationary point $\bm{s}^*$, we still can not use $\bm{s}^*$ to directly represent the steady system state, as there may be multiple stationary points or the state $\bm{s}(t)$ may not converge to $\bm{s}^*$. This requires us to further address the problems of unique  stationary point and convergence of  state $\bm{s}(t)$.   

\subsubsection{Uniqueness and convergence}
Due to the graph structure, mean field model also depends on the degree distribution, which makes it difficult to characterize the unique stationary point and state convergence. To circumvent this problem, we will show the uniqueness and convergence issues  alternatively. 

\textbf{Coordinate-wise dominance}. The state evolution of  $s_{k,i}(t)$ is identified by the ODE of Eq.~\eqref{Eq:ODE-s}. To obtain the state convergence, we have to figure out how the initial values $\bm{s}(0)$ would influence the state $\bm{s}(t)$ at later time $t$. Define coordinate-wise dominance $\bm{s} \succeq \bm{\hat{s}}$ if $s_{k,i} \ge  \hat{s}_{k,i}, \forall k \in \mathcal{K}, i \ge 0$. The lemma below states that the dominance at any time $t$ is consistent with that of the initial values.

\begin{lem} \label{lem:dominance}
Let $\bm{s}(t)$ and $\bm{\hat{s}}(t)$ be the solutions to the ODE system of Eq.~\eqref{Eq:ODE-s} at time $t$ with the initial values being $\bm{s}(0)$ and $\bm{\hat{s}}(0)$, respectively. If $\bm{s}(0) \succeq \bm{\hat{s}}(0)$, then $\bm{s}(t) \succeq \bm{\hat{s}}(t)$.
\end{lem}

See Appendix~\ref{App:dominance} for the proof. With the dominance consistency, we demonstrate that every trajectory of the state converges to the stationary point in an appropriate metric.

\textbf{Exponential convergence rate}. To show convergence, we need to find a Lyapunov function $\phi(\mathbf{s})$ which satisfies: 1)  $\phi(\mathbf{s})$ relates to the distance between $\bm{s}$ and $\bm{s}^*$; 2) $\phi(\mathbf{s})$ is strictly decreasing, except at $\bm{s}^*$. Here, $\phi(\mathbf{s})$ is constructed as:
\begin{equation}
\phi(\mathbf{s}) = \min_{\bm{s}^* \in \mathcal{S}^*} \sum_{i \ge 0} \frac{|\sum_{k \in \mathcal{K}} p(k)(s_{k,i}-s_{k,i}^*)|}{2^i},
\end{equation}
where $\mathcal{S}^*$ is the stationary point set. To simplify $\phi(\mathbf{s})$, we denote $s_i = \sum_{k \in \mathcal{K}} p(k) s_{k,i}$ and $s_{(k),i} = \sum_{k \in \mathcal{K}} p(k) k s_{k,i}$. In line with Eq.~\eqref{Eq:ODE-s}, the evolution of $s_i $ is specified as:
\begin{equation} \label{Eq:ODE-sk}
\dot{s}_i = \frac{x_c \lambda}{\overline{k}} \left(s_{i-1}s_{(k),i-1} - s_{i}s_{(k),i}\right) - \mu(s_i -s_{i+1}).
\end{equation}
Using $s_i$, we have $\phi(\mathbf{s}) =  \min_{\bm{s}^* \in \mathcal{S}^*} \sum_{i \ge 0} \frac{|s_i -s_i^*|}{2^i}$.  The convergence of state $\bm{s}$ is derived by showing $\phi(\mathbf{s}) \rightarrow 0$.

\begin{lem} \label{lem:convergence}
If initial points $\bm{s}(0) \succeq \bm{s}^*, \forall \bm{s}^* \in \mathcal{S}^*$ or $\bm{s}^* \succeq \bm{s}(0), \forall \bm{s}^* \in \mathcal{S}^*$,  $\bm{s}(t)$ converges to $\mathcal{S}^*$ with exponential rate.
\end{lem}
Please refer to Appendix~\ref{App:convergence} for the detailed proof. The exponential convergence rate in Lemma~\ref{lem:convergence} is derived under certain initial conditions. However, if there is only one stationary point, global  convergence from any initial point will be naturally obtained. This is because when there is a unique stationary point $\bm{s}^*$, $\bm{s}(t)$ always converges to $\bm{s}^*$ if $\bm{s}(0) \succeq \bm{s}^*$ or $\bm{s}^* \succeq \bm{s}(0)$. The dominance consistency in Lemma~\ref{lem:dominance} then ensures the convergence of $\bm{s}(t)$ to $\bm{s}^*$  from other initial values. 

\textbf{Unique stationary point}. Now we  show that our mean field model has a  unique stationary point, which will be proved by combining the dominance consistency property and the exponential convergence result. 

\begin{thm} \label{thm:uniqueness}
There is a unique stationary point $\bm{s}^*$ for the mean field model.
\end{thm}

See Appendix~\ref{App:uniqueness} for the proof. Lemma~\ref{lem:convergence} and Theorem~\ref{thm:uniqueness} imply the result as stated in the following theorem.
\begin{thm} \label{thm:convergence}
The state evolution of D2D collaboration globally converges to a unique stationary point with exponential rate.
\end{thm}

\subsection{Influence of Heterogeneous Degrees}
In minimizing the system cost, one constraint is the collaboration fairness in Eq.~\eqref{Eq:fairCon} pertaining to heterogeneous degrees. Concretely, a user with larger degree receives more tasks in D2D collaboration, meanwhile  having a higher probability to forward tasks to its neighbors. This effect is presented below.


\begin{thm} \label{thm:heter_degree}
At the stationary point $\bm{s}^*$, larger-degree users tend to have heavier workloads. In other words, $s^*_{k,i} \ge s^*_{k',i}, \forall i \ge 0$ if $k>k'$.
\end{thm}

\begin{figure}[htp]
  \centering
  \includegraphics[width=0.8\columnwidth]{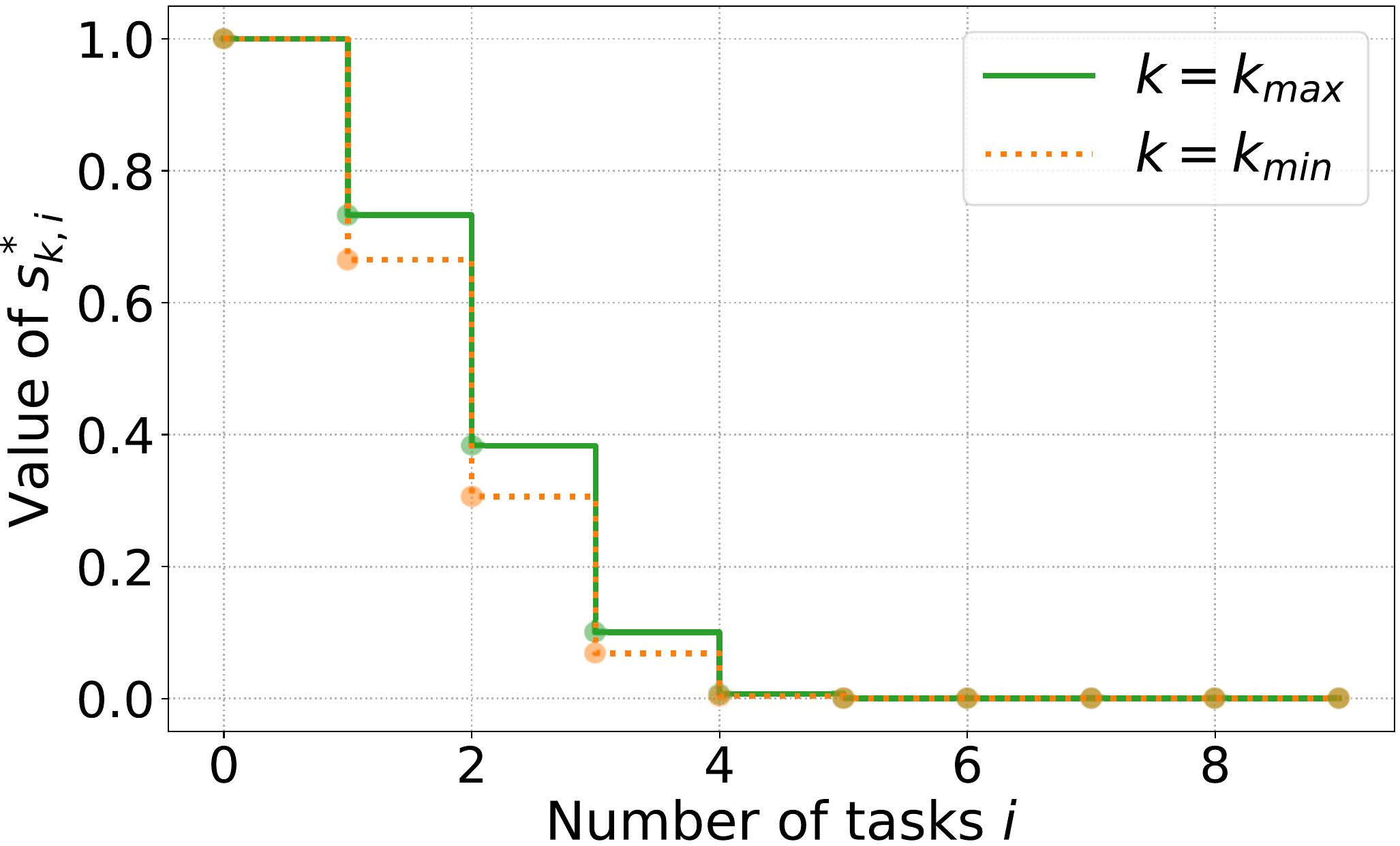}
  \caption{Illustration of workload distribution.}\label{Fig:ski}
  \vspace{-15pt}
\end{figure}

See Appendix~\ref{App:heter_degree} for the proof. This theorem reveals that users with heterogeneous degrees have \emph{uneven workloads} because  larger-degree users generally have higher contributions in D2D collaboration. Note that $s_{k,1}^*$ indicates the busy probability of a $k$-degree user with processing tasks. To avoid small-degree users free-ride large-degree neighbors, we bound  the gap between $\max\{s_{k,1}^*\}$ and $\min\{s_{k,1}^*\}$ in the system cost minimization so as to ensure the collaboration fairness, where  $\max\{s_{k,1}^*\}$ and $\min\{s_{k,1}^*\}$ now become  $s_{k_{\max},i}^*$ and $s_{k_{\min},i}^*$, respectively,  from Theorem~\ref{thm:heter_degree}. Fig.~\ref{Fig:ski} compares typical values of $s_{k_{\max},i}^*$ and $s_{k_{\min},i}^*$  for  some basic understanding.

\subsection{Convergence to Mean Field Model}
The mean field model  makes use of a deterministic ODE system to analyze D2D collaboration in a stochastic MEC system. In the following, we show that the original stochastic $N$-user system will converge to the deterministic mean field model when $N$ is large, namely the ODE system of Eq.~\eqref{Eq:ODE-s} is accurate in describing the state evolution.

Consistent with Eq.~\eqref{Eq:Por-s},  we denote $\bm{s}^{(N)}(t)$ as the state of  $N$-user system. Our goal is to  demonstrate that $\bm{s}^{(N)}(t) \rightarrow \bm{s}(t)$ when $N \rightarrow \infty$. To arrive at this conclusion, we first prove the drift function $\bm{F}(\bm{s})$ in Eq.~\eqref{Eq:drift} is  Lipschitz continuous.

\begin{lem} \label{lem:drift}
The drift function $\bm{F}(\bm{s})$ is $||\cdot||_{\infty}$-Lipschitz as there is a constant $C=3x_c \lambda \big(1+ \frac{k_{\max}}{\overline{k}}\big) +2 \mu$ such that for any $\bm{s}, \bm{\hat{s}}$:
\begin{equation} \label{Eq:Lipschitz}
||\bm{F}(\bm{s}) - \bm{F}(\bm{\hat{s}})||_{\infty} \le C ||\bm{s}-\bm{\hat{s}}||_{\infty}.
\end{equation}
\end{lem}
See Appendix~\ref{App:drift} for the proof. With Lemma~\ref{lem:drift}, we can claim the convergence of  $N$-user system to the mean field model based on the  Kurtz's theorem~\cite{kurtz1981approximation}.
\begin{thm} \label{thm:efficacy}
Fix a time $t^*$. When the number of users $N \rightarrow \infty$, the state $\bm{s}^{(N)}(t)$ converges to $\bm{s}(t)$ in the mean field model of Eq.~\eqref{Eq:ODE-s} if they start from the same initial points.
\begin{equation} \label{Eq:Ninfty}
\lim_{N \rightarrow \infty} \sup_{t \in [0,t^*]} ||\bm{s}^{(N)}(t) - \bm{s}(t)||_{\infty} = 0, \mathrm{a.s.}
\end{equation}
\end{thm}

Please refer to Appendix~\ref{App:efficacy} for the proof. Theorem~\ref{thm:efficacy} ensures that  the mean field model is effective for large population $N$. On this basis, one can obtain  the existence and uniqueness of  stationary point to characterize  the steady  state of MEC system and show the power of D2D collaboration.

\subsection{Discussion of Stationary Point}
\subsubsection{Relation to classical Po2} The mean field models on static and dynamic graphs are unified by the ODE system of Eq.~\eqref{Eq:ODE-s}. When users have a homogeneous degree, Eq.~\eqref{Eq:ODE-s} boils down to the classical Po2 of Eq.~\eqref{Eq:ODE-Spe}, which is independent of the degree and has a closed-form stationary point~\cite{mitzenmacher2001the}:
\begin{equation} \label{Eq:Tra_Po2}
\pi_{i}^* = \Bigl(\frac{x_c \lambda}{\mu} \Bigr)^{2^i -1}.
\end{equation}
For a more general graph of heterogeneous degrees, an explicit expression for stationary point $\bm{s}^*$ is not available. Nevertheless, we can use the traditional Po2 as a  bound for $\bm{s}^*$. Define two ratios:
\begin{equation} \label{Eq:ratio}
\delta_1 = \frac{k_{\max}}{\overline{k}},~\delta_2 = \frac{k_{\min}}{\overline{k}}.
\end{equation}
Here, $\delta_1 >1$ and $\delta_2 <1$. Also, let $\frac{1+\delta_1}{2} \frac{x_c \lambda}{\mu} <1$ to guarantee the system stability.

\begin{cor} \label{cor:bound}
For the mean field model of Eq.~\eqref{Eq:ODE-s}, if $\frac{1+\delta_1}{2} \frac{x_c \lambda}{\mu} <1$, then $s^*_{k,i}, \forall k, i$ has an upper bound:
\begin{equation} \label{Eq:upper}
s^*_{k,i} \le  \Bigl(\frac{1+\delta_1}{2} \frac{x_c \lambda}{\mu} \Bigr)^{2^i -1},
\end{equation}
and a lower bound:
\begin{equation} \label{Eq:lower}
s^*_{k,i} \ge  \Bigl(\frac{1+\delta_2}{2} \frac{x_c \lambda}{\mu} \Bigr)^{2^i -1}.
\end{equation}
\end{cor}

See Appendix~\ref{App:bound} for the proof.  According to this corollary,  if the degree distribution is slightly heterogeneous, i.e., both the ratios $\delta_1$ and $\delta_2$ in Eq.~\eqref{Eq:ratio} are close to 1, the gap between upper and lower bounds will be small, thereby leading to an accurate estimation of $s_{k,i}^*$ with closed-form expressions. 

\subsubsection{Busy probability} The probability that a user is busy with processing task, $s^*_1 = \sum_{k \in \mathcal{K}} p(k)s_{k,1}^*$,  is critical in computing the system cost in Eq.~\eqref{Eq:cost}. The corollary below provides the value of $s^*_1$ with the proof presented in Appendix~\ref{App:busy}.
\begin{cor} \label{cor:busy}
The  busy probability is $s^*_1  = \frac{x_c \lambda}{\mu}$.
\end{cor}


\subsubsection{Workload distribution relation} In the end of this section, we illustrate  the relation between $s_{i}^* $ and $s_{i-1}^*$.  Based on Eq.~\eqref{Eq:Tra_Po2},  traditional Po2 satisfies $\pi_{i}^* =  \frac{x_c \lambda}{\mu} \left( \pi^*_{i-1} \right)^2$. As for our mean field model on graph, similar conclusion is attained. Specifically, from  Eq.~\eqref{Eq:ODE-sk}, we know that the stationary point satisfies the following condition:
\begin{equation} \label{Eq:inv_sk}
\frac{x_c \lambda}{\overline{k}} \left(s^*_{i-1} s_{(k),i-1}^* -s^*_{i} s_{(k),i}^* \right) - \mu(s^*_i - s^*_{i+1})=0.
\end{equation}

\begin{cor} \label{cor:relation}
For any $i \ge 1$, we have $s_{i}^* =  \frac{x_c \lambda}{\overline{k} \mu} s^*_{i-1} s^*_{(k),i-1}$. 
\end{cor}

See Appendix~\ref{App:relation} for the proof. Compared to the classical Po2, the graph structure causes the difference between $\pi_{i}^* =  \frac{x_c \lambda}{\mu}  \pi^*_{i-1}  \times \pi^*_{i-1}$ and $s_{i}^* =  \frac{x_c \lambda}{\mu} s^*_{i-1} \times \frac{s^*_{(k),i-1}}{\overline{k}}$.


\section{Online Offloading and Pricing Scheme} \label{Sec:price}
D2D collaboration can effectively reduce workloads of users. However, due to constrained resources of mobile devices, offloading a portion of tasks to a more powerful edge server is still essential to further mitigate task execution delay. Along with the offloading, there is a  price charged by the edge server, so that users have to balance how many tasks should be offloaded and how many should be processed collaboratively. Task offloading between mobile users and edge server is modeled as a Stackelberg game, where the server is the leader in setting a service  price which remains fixed for a long time,  and users are followers in deciding the offloading probability. 

%

\subsection{Lyapunov Optimization}
With \emph{currently available information}, the server is interested in maximizing its long-term utility subject to the overloaded constraint by setting a proper price $p[n]$ in each time slot, as shown in Eq.~\eqref{Eq:utility_max}. Meanwhile, mobile users aim to reduce their system cost while maintaining satisfactory task delay and collaboration fairness through determining the offloading probability $x[n]$, which is described in Eq.~\eqref{Eq:cost_min}. To achieve these two goals, \emph{Lyapunov optimization is leveraged to maximize the long-term utility online, with each time slot corresponding to a Stackelberg game to  minimize the system cost offline.}

\subsubsection{Optimal task offloading}
At the beginning of time slot $n$, assume the edge server has declared a price $p[n]$. Users then collectively determine the offloading decision $x[n]$ to minimize  their average system cost $c[n]$. Combining  Eq.~\eqref{Eq:cost} and the busy probability in Corollary~\ref{cor:busy}, we can rewrite  $c[n]$ as:
\begin{equation} \label{Eq:re-cost}
c[n] =  x[n] \lambda p[n] + (1-x[n])\lambda  \frac{\rho_c^m}{\mu^2} +  x[n] \lambda \rho_t^m \frac{B}{r}. 
\end{equation}


\textbf{Critical points}. Task delay $d(x[n])$ in Eq.~\eqref{Eq:delay} is composed of two parts. The first part is the transmission delay and completion time of task offloading, that is $d_o = x[n] \big(\frac{B}{r}+\frac{1}{\gamma}\big)$. The second part corresponds to the sojourn time of D2D collaboration $d_q=\frac{\sum_{i}\sum_{k}p(k)s_{k,i}^*}{\lambda}$.  If  $x[n]$ increases from $0$ to $1$, $d_o$  will monotonically increase whereas  $d_q$ will monotonically decrease. Hence, there exist a lower bound $x_l^{*}$ and an upper bound $x_u^{*}$ such that when $x[n] \in [x_l^{*},x_u^{*}]$ the delay constraint is fulfilled, where critical points $x_l^{*},x_u^{*}$ satisfy:
\begin{equation} \label{Eq:critical1}
d\left(x[n]=x_l^{*}\right) =d\left(x[n]=x_u^{*}\right) = \overline{d}.
\end{equation}
Note that  the stationary point $\bm{s}^*$ is dependent on $x[n]$. We will compute each $s_{k,i}^*$  numerically given  the value of $x[n]$, say $x[n] = x_l^{*}, x_u^{*}$,  since there is no closed-form solution.

\begin{figure}[htp]
  \centering
  \includegraphics[width=0.8\columnwidth]{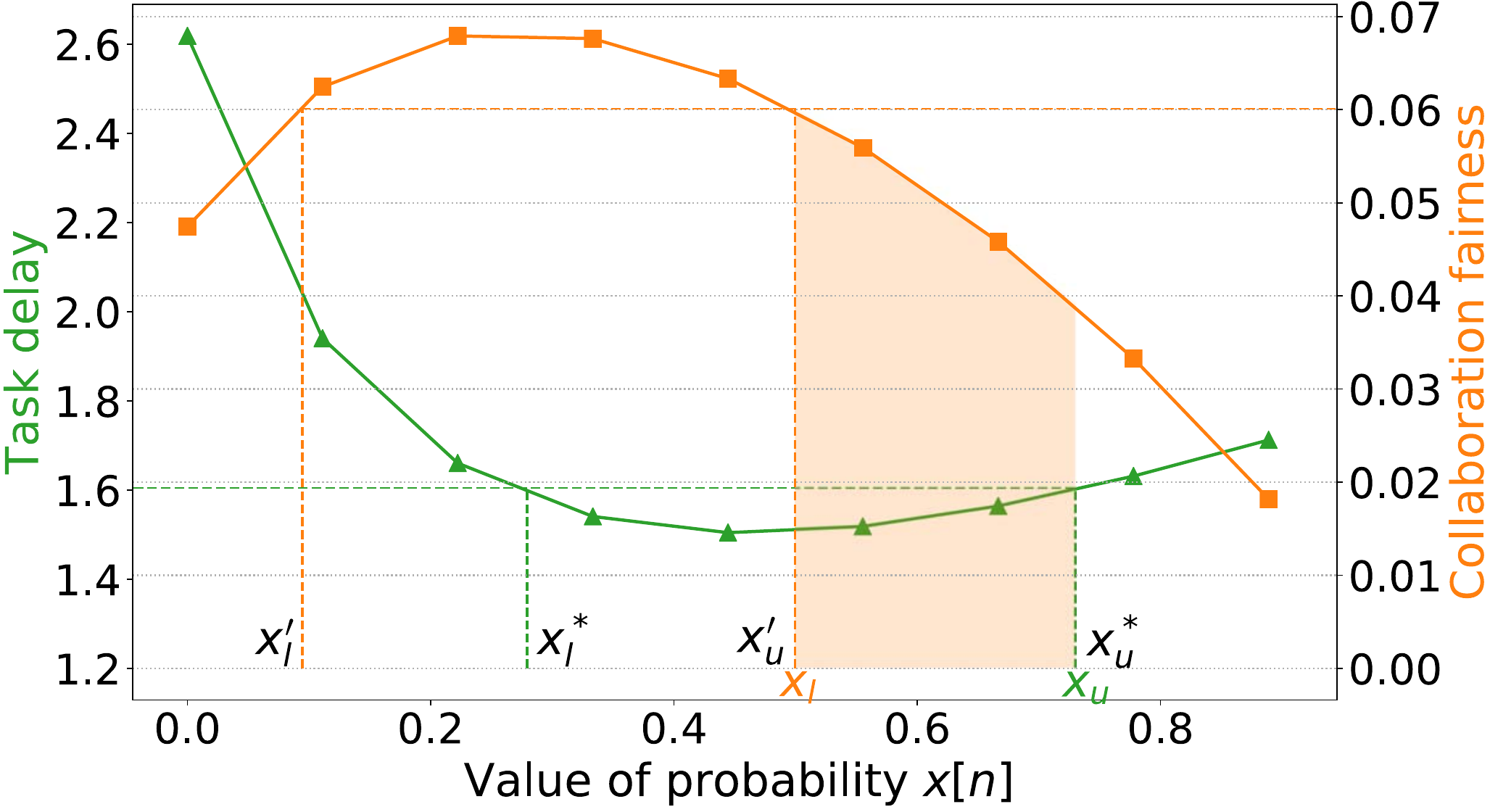}
  \caption{Feasible region for offloading decision.}\label{Fig:constraint}
  \vspace{-6pt}
\end{figure}

Let us discuss the collaboration fairness constraint, which is $s_{k_{\max},1}^* - s_{k_{\min},1}^* \le \overline{s}$ according to Theorem~\ref{thm:heter_degree}. The trend of the gap $s_{k_{\max},1}^* - s_{k_{\min},1}^*$ over $x[n]$ is not obvious. Nevertheless, if $x[n]$ approaches $1$, i.e., users offload all tasks to the edge server,  both $s^*_{k_{\max},1}$ and $s^*_{k_{\min},1}$ will be $0$, and then their gap will be $0$. When we push $x[n]$ approaching $0$, that is users do not offload but only collaborate,  all users will be heavily loaded, and hence $s^*_{k_{\max},1}$ and $s^*_{k_{\min},1}$ will be close to $\frac{\lambda}{\mu}$, with the gap being very small.  As a result, we can characterize two feasible regions for $x[n]$: $[0,x'_l] \cup [x'_u,1]$.  If jointly considering the delay and fairness constraints,  Fig.~\ref{Fig:constraint} provides a typical illustration of the feasible region for the offloading decision $x[n]$. In this paper, we assume that $[x_l^{*},x_u^{*}] \cap ([0,x'_l] \cup  [x'_u,1]) \ne \emptyset$, namely $x[n]$ has a feasible solution.  Additionally, for any feasible region, we denote $x_l$ and $x_u$ as its lower and upper boundary points, respectively, which hinge on  the values of  $x_l^{*},x_u^{*}, x'_l,x'_u$, as also displayed in Fig.~\ref{Fig:constraint}.

\textbf{Threshold based offloading decision}. To minimize the system cost in Eq.~\eqref{Eq:re-cost}, provided the price $p[n]$,  the optimal probability $x[n]$ for users is specified by a \emph{threshold based decision}:
\begin{equation} \label{Eq:xsolution}
x[n]=
\left \{
\begin{aligned}
&x_u~&& \mathrm{if}~\frac{\rho_c^m}{\mu^2} \ge p[n]+\rho_t^m \frac{B}{r},\\
&x_l &&\mathrm{otherwise}.
\end{aligned}
\right.
\end{equation}
The intuition behind Eq.~\eqref{Eq:xsolution} is now explained. If the server sets an excessively high price such that the cost of offloading  task is greater than the cost of processing task collaboratively, users will offload as few tasks as possible to reduce their system cost, and vice versa. With this reacted offloading decision $x[n]$,  the edge server in turn determines the optimal price $p[n]$ in each time slot to maximize its service utility. Note that the critical points $x_l,x_u$ are altered by $p[n]$ in that the price will affect the offloading decision. Recall from the overloaded  constraint  in Eq.~\eqref{Eq:overload},  it implies that  $\overline{x} \in [x_l\lambda, x_u\lambda]$.

\subsubsection{Dynamic service pricing}
The server will judiciously choose the price over time to maximize its long-term utility, which follows an online Lyapunov optimization framework. 

\textbf{Drift-minus-utility}. In view of the overloaded constraint in utility maximization, we define a virtual queue $X[n]$ for the edge server which buffers the virtual amounts of offloaded tasks. Here, we use the prefix ``virtual'' to denote that  tasks are not actually offloaded from users, but rather, to reflect the requirement of the overloaded constraint. Consistent with this queue definition, tasks will enter into the queue with arrival rate $x[n]\lambda$ where $x[n]$ is users' offloading probability, and will leave the queue with departure rate $\overline{x}$. Therefore, we have the following dynamic equation for the virtual queue $X[n]$:
\begin{equation} \label{Eq:queue}
X[n+1] = \max (X[n]+x[n] \lambda -\overline{x},0).
\end{equation}
Based on Eq.~\eqref{Eq:queue}, we construct Lyapunov function as $\frac{1}{2}X^2[n]$, and compute Lyapunov drift which basically is the change of Lyapunov function from one time slot to the next:
\begin{equation} \label{Eq:Lya_drift}
\Delta (X[n]) =\mathbb{E}\Bigl[\frac{1}{2}X^2[n+1] - \frac{1}{2}X^2[n] \big| X[n] \Bigr].
\end{equation}
The expectation is taken over  the randomness in task generation, offloading decision and pricing scheme. By minimizing Lyapunov drift $\Delta (X[n])$, one  can drive the queue backlog to a small value so as to maintain $X[n]$ rate stable, $\lim_{n \rightarrow \infty} \frac{X[n]}{n}=0$, with probability $1$. From the queue stability theorem~\cite{neely2010stochastic},  a queue $X[n]$ is stable if and only if the arrival rate is no larger than the departure rate, i.e., $\lim_{T \rightarrow \infty} \frac{1}{T} \sum \nolimits_{n=0}^{T-1} \mathbb{E}[x[n]\lambda] \le \overline{x}$, and hence the overloaded constraint is satisfied.  Furthermore, we define drift-minus-utility $\Delta (X[n]) - V  \mathbb{E}[u[n]|X[n]]$, where $V$ is the importance weight on the utility term. Minimizing  drift-minus-utility will \emph{simultaneously push the queue backlog  to a small value and maximize the  utility} as well~\cite{neely2010stochastic}. 

\textbf{Bound of drift-minus-utility}. In time slot $n$, the queue $X[n]$ is known in advance. Besides, the offloading decision $x[n]$ is given  in Eq.~\eqref{Eq:xsolution}, which is also expressed as $x(p[n])$ to explicitly indicate its dependence on the price $p[n]$. 

\begin{lem} \label{lem:drift_bound}
Drift-minus-utility $\Delta (X[n]) - V \mathbb{E}[u[n]|X[n]]$ satisfies the following condition:
\begin{equation} \label{Eq:drift_bound}
\begin{aligned}
&\Delta (X[n]) - V \mathbb{E}[u[n]|X[n]]  \le  \mathbb{E}\big[X[n](x(p[n])\lambda - \overline{x})|X[n]\big] \\
& - \mathbb{E}\Bigl[\big(V x(p[n]) \lambda p[n] -V x(p[n]) \lambda \frac{\rho_c^s}{\gamma} \big)\big|X[n] \Bigr] + D,
\end{aligned}
\end{equation}
where $D = \max(\frac{1}{2} \left(\lambda-\overline{x})^2, \frac{1}{2} \overline{x}^2 \right)$.
\end{lem}

See Appendix~\ref{App:drift_bound} for the proof. Lemma~\ref{lem:drift_bound} allows us to use a \emph{simple-form bound} rather than the original complex drift-minus-utility in deriving the optimal price. As $X[n]$ is a priori knowledge in each time slot $n$ and $D$ is a constant, the remaining terms of Eq.~\eqref{Eq:drift_bound} precisely correspond to the overloaded constraint and the service utility, respectively.

\textbf{Optimal service price}. In every time slot $n$, we minimize  the bound of  drift-minus-utility with $x(p[n])$ in Eq.~\eqref{Eq:xsolution}:
\begin{subequations}  \label{Eq:opt_slot}
\begin{align}
\min~&\scalebox{0.91}{$X[n](x(p[n])\lambda - \overline{x}) -Vx(p[n]) \lambda p[n]  +V x(p[n]) \lambda \frac{\rho_c^s}{\gamma} $}  \tag{\ref{Eq:opt_slot}}\\
\mathrm{s.t.}~~& p[n] \in (0,p_u]  \label{Eq:accept1}.
\end{align}
\end{subequations}
First, suppose that $p[n] \le \frac{\rho_c^m}{\mu^2} - \rho_t^m \frac{B}{r}$, then $x(p[n]) = x_u$, thus the objective of Eq.~\eqref{Eq:opt_slot} is $u_1(p[n])= X[n] \left(x_u \lambda - \overline{x}\right) -V x_u \lambda p[n]  +V  x_u \lambda \frac{\rho_c^s}{\gamma}$. The price $p[n]$ should be $\frac{\rho_c^m}{\mu^2} - \rho_t^m \frac{B}{r}$ in order to minimize $u_1(p[n])$.  Following this line, assume that $p[n] > \frac{\rho_c^m}{\mu^2} - \rho_t^m \frac{B}{r}$, and the objective becomes $u_2(p[n]) = X[n](x_l \lambda - \overline{x}) -V x_l \lambda p[n]  +V x_l \lambda \frac{\rho_c^s}{\gamma}$, thereby $p[n]$ ought to be $p_u$. In general, we compare the two values $u_1(p[n]= \frac{\rho_c^m}{\mu^2} - \rho_t^m \frac{B}{r}) =X[n]( x_u \lambda - \overline{x}) -V x_u \lambda (\frac{\rho_c^m}{\mu^2} - \rho_t^m \frac{B}{r})  +V  x_u \lambda \frac{\rho_c^s}{\gamma}$ and $u_2(p[n]= p_u) = X[n](x_l \lambda - \overline{x}) -V x_l \lambda p_u  +V x_l \lambda \frac{\rho_c^s}{\gamma}$ to determine $p[n]$. Let $X^* =  \frac{V x_u(\frac{\rho_c^m}{\mu^2} - \rho_t^m \frac{B}{r})-Vx_l p_u}{ x_u-x_l} -V\frac{\rho_c^s}{\gamma}$, then:
\begin{equation} \label{Eq:opt_price}
p[n]=
\left \{
\begin{aligned}
& \frac{\rho_c^m}{\mu^2} - \rho_t^m \frac{B}{r}~&&\mathrm{if}~X[n] \le X^*,\\
&p_u  &&\mathrm{otherwise}.
\end{aligned}
\right.
\end{equation}
After setting price $p[n]$, offloading decision $x[n]$ is made based on Eq.~\eqref{Eq:xsolution}, and queue $X[n]$ is updated following Eq.~\eqref{Eq:queue}.

 \subsection{Performance of  Offloading and Pricing Scheme}
Now we provide the performance analysis of our task offloading and service pricing. In particular, we demonstrate that an asymptotically optimal utility is obtained and the queue backlog $X[n]$ has a constant upper bound.
 
 \begin{thm} \label{thm:Lyapunov}
 Suppose the initial queue backlog $X[0] = 0$. For any importance weight $V>0$, the proposed task offloading and service pricing satisfy the following properties.
 
 a) The queue backlog in any time slot $n$ is bounded:
 \begin{equation} \label{Eq:Xn_bound}
X[n] \le  \frac{V x_u(\frac{\rho_c^m}{\mu^2} - \rho_t^m \frac{B}{r})-Vx_l p_u}{ x_u-x_l} -V\frac{\rho_c^s}{\gamma} +  x_u \lambda - \overline{x}.
 \end{equation}
 
 b) Denote $u^*$ as the optimal time average utility for Eq.~\eqref{Eq:utility_max}, then the achieved utility satisfies:
 \begin{equation} \label{Eq:Lyapunovu}
\lim_{T \rightarrow \infty} \frac{1}{T} \sum_{n=0}^{T-1} \mathbb{E}[u[n]]  \ge u^* - \frac{D}{V}.
 \end{equation}
 \end{thm}

Please refer to Appendix~\ref{App:Lyapunov} for the detailed proof. Theorem~\ref{thm:Lyapunov} unveils an $(O(V),O(1/V))$ tradeoff between the queue backlog and the service utility. Specifically, as the importance weight $V$ increases, the queue backlog also increases as fast as the order of  $O(V)$, while the time average utility approaches the theoretical optimum within an $O(1/V)$ gap. In addition, because $X[n]$ is upper bounded by a finite value given in Eq.~\eqref{Eq:Xn_bound}, and then $X[n]$ is rate stable, that is the overloaded constraint will hold asymptotically.

\section{Performance Evaluation} \label{Sec:performance}
In this section, we carry out evaluations to illustrate D2D collaboration among users and  task offloading to edge server, especially to evaluate the mean field model on graph and Lyapunov optimization based  offloading and pricing scheme. 
%
\subsection{Parameter Setting} \label{Sec:performance}
In line with the real measurements~\cite{kwak2015dream}, we set the average  service time of a task to $1000$ Megacycles which is normalized to $1$ as stated in the system model, and the average data size $B$ to $2000$ KB. Service rates of a mobile device and the edge server are $1$ GHz and $5$ GHz, respectively, thus $\mu = \frac{10^9}{1000*10^6}=1$ and $\gamma = \frac{5*10^9}{1000*10^6}=5$ accordingly. Moreover, the typical real-world data rate of $4$G cellular uplink is around $r=10$ Mbps~\cite{4G4U20194G}.

For D2D collaboration, task generation rate $\lambda$ is set to 0.9 for modeling a heavy workload situation. User's (expected) degree  in   (dynamic) static graph is uniformly distributed  in the set  $\mathcal{K}=\{6,7,8,9\}$, that is $p(k) =\frac{1}{4}, \forall k \in \mathcal{K}$ and $k_{\min}=6, k_{\max} = 9$. Besides, user's realized degree given its expected degree obeys a Poisson distribution for dynamic graph~\cite{lang2018analytic}, and the graph structure varies over time with rate $1$.

Regarding  task offloading, let $\overline{d} = 1.6$ for the delay constraint, and $\overline{s} = 0.06$ for the collaboration fairness constraint. The per energy costs are $(\rho^m_c,\rho^m_t,\rho^s_c) = (0.9,0.3,1)$, which correspond to $900$ mW, $300$ mW for processing and transmitting tasks in mobile devices, and  $1000$ mW for processing tasks in the edge server. The overloaded threshold $\overline{x}$ is set to $0.6$, and  the highest acceptable price $p_u$ is assigned to $0.5$.

\begin{figure*}[t]
\centering
\begin{tabular}{cc}
    \begin{minipage}[t]{1.6in}
    \includegraphics[height=0.9in,width=1\columnwidth]{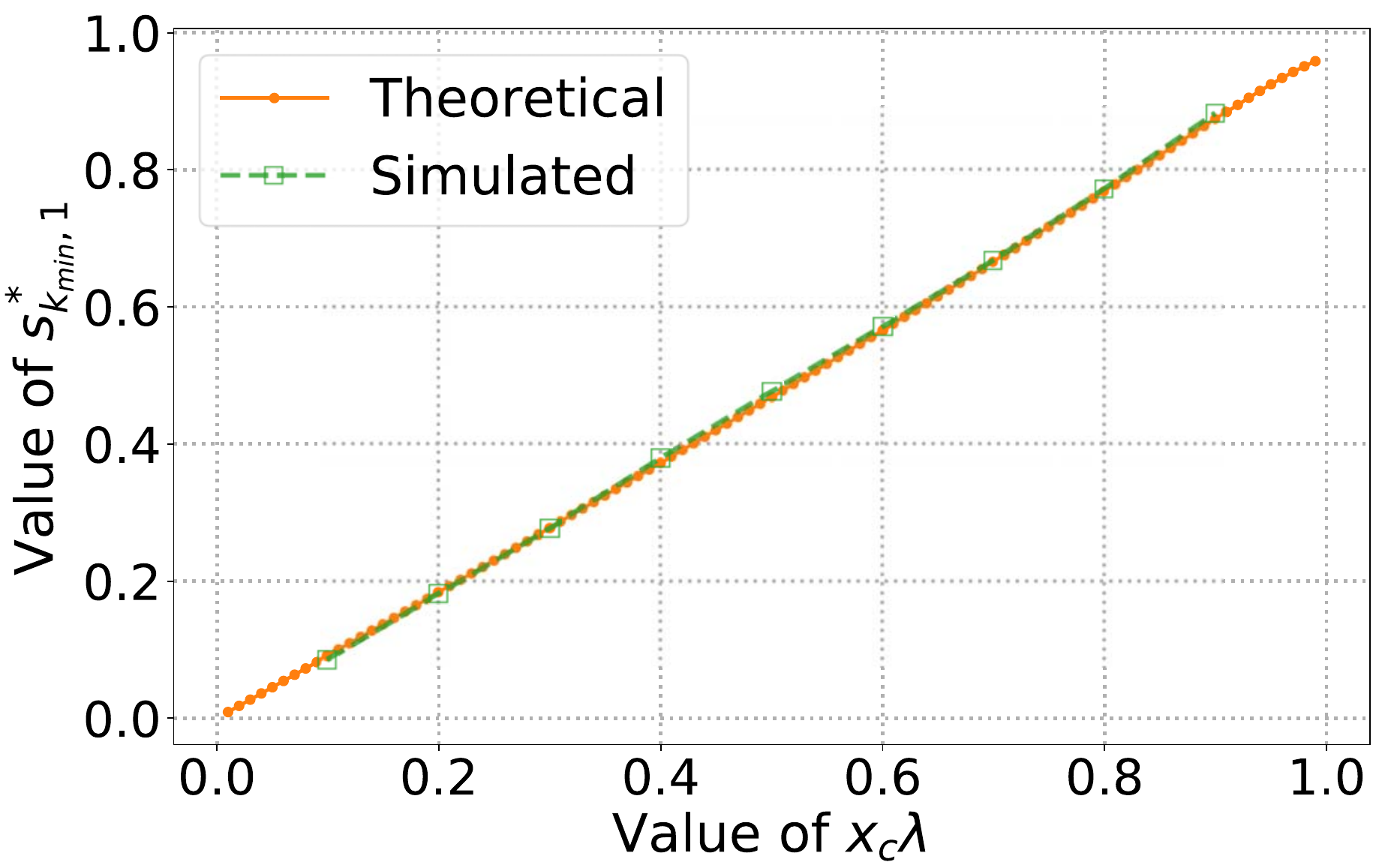}
    \caption{Stationary point on static graph.}
    \label{Fig:MF_static}
    \end{minipage}

    \begin{minipage}[t]{1.6in}
    \includegraphics[height=0.9in,width=1\columnwidth]{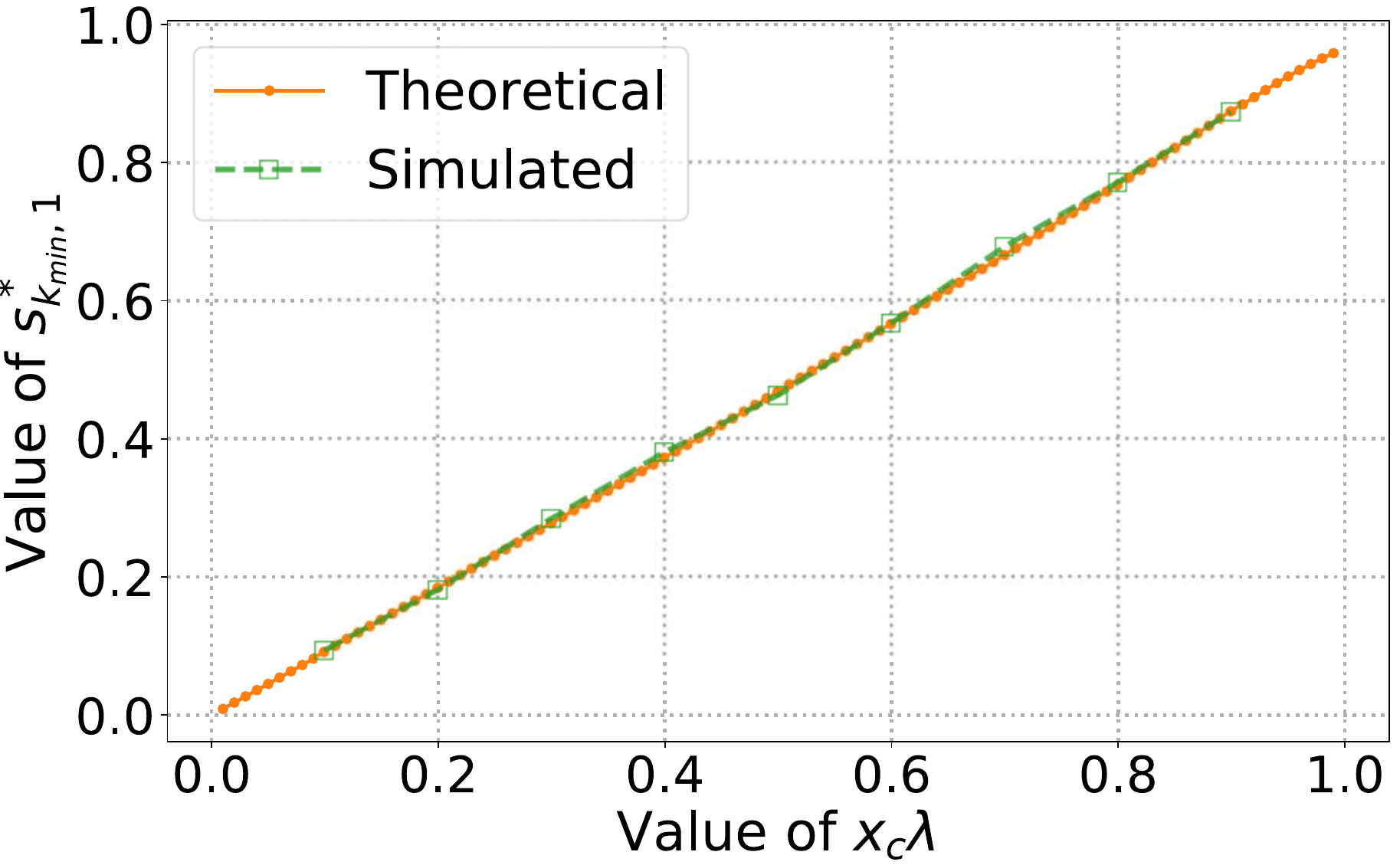}
    \caption{Stationary point on dynamic graph.}
    \label{Fig:MF_dynamic}
    \end{minipage}

    \begin{minipage}[t]{1.6in}
    \includegraphics[height=0.9in,width=1\columnwidth]{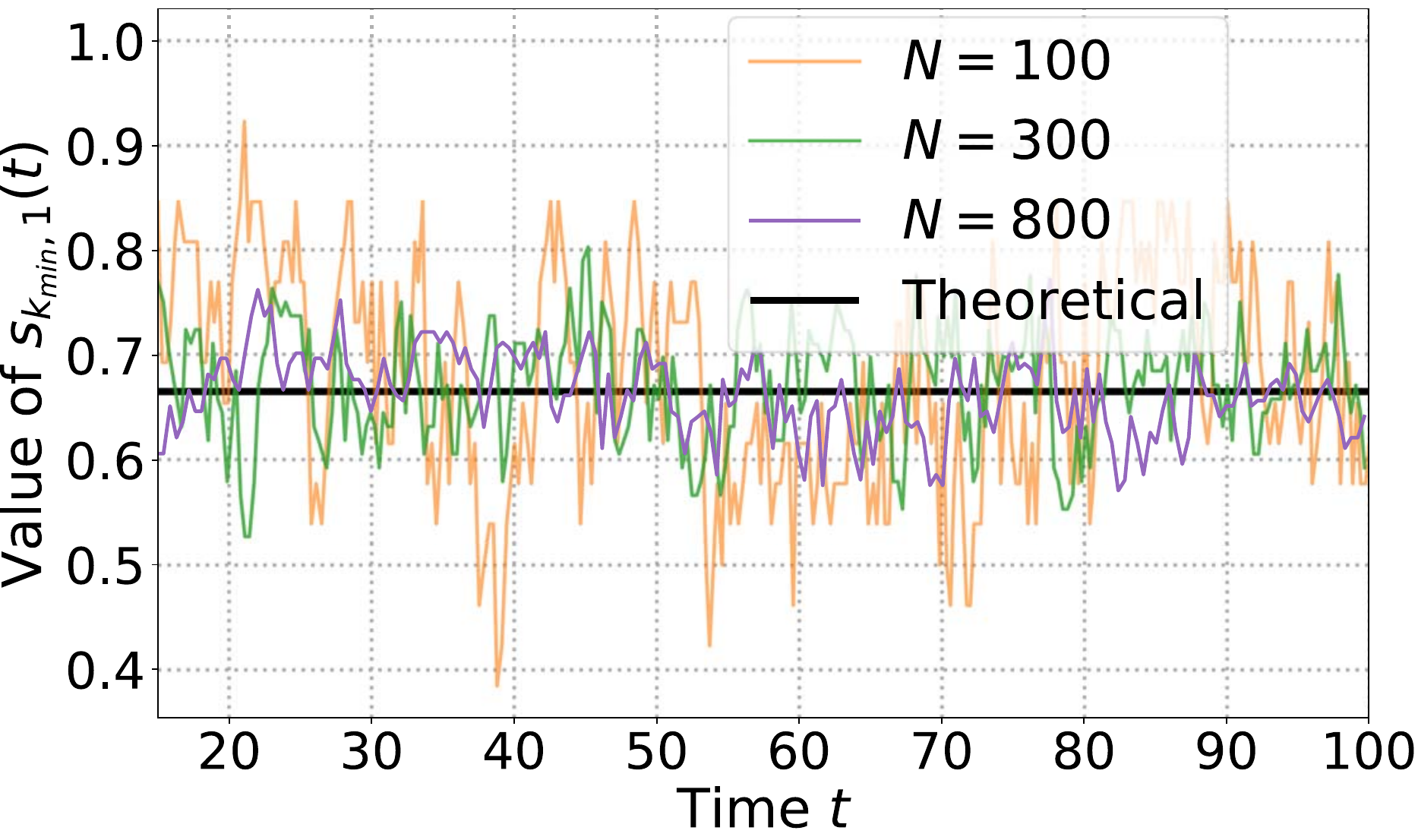}
    \caption{Convergence to mean field model on static graph.}
    \label{Fig:MF_static_N}
    \end{minipage}

    \begin{minipage}[t]{1.6in}
    \includegraphics[height=0.9in,width=1\columnwidth]{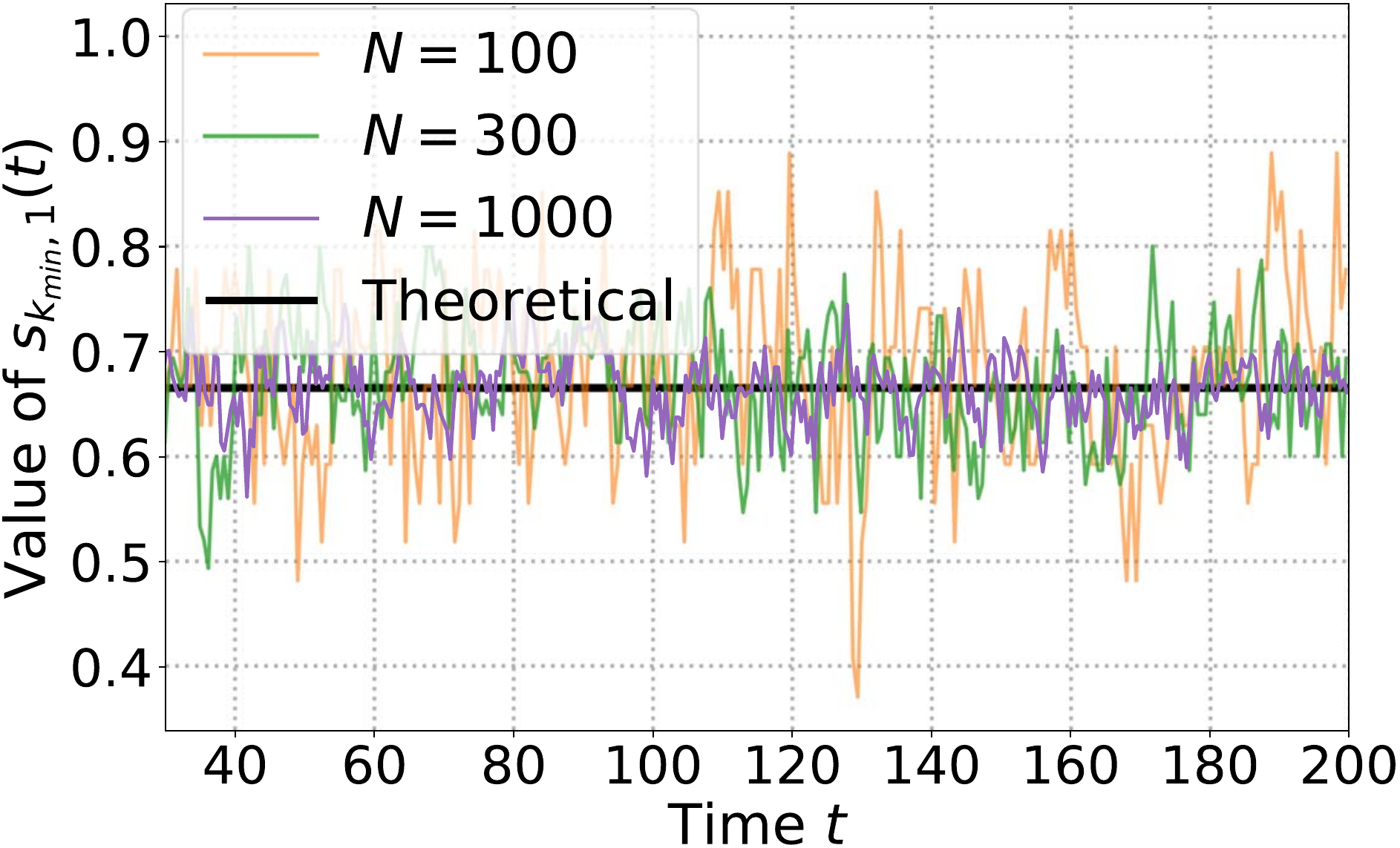}
    \caption{Convergence to mean field model on dynamic graph.}
    \label{Fig:MF_dynamic_N}
    \end{minipage}

\end{tabular}
\vspace{-10pt}
\end{figure*}

\begin{figure*}[t]
\centering
\begin{tabular}{cc}
    \begin{minipage}[t]{1.6in}
    \includegraphics[height=0.9in,width=1\columnwidth]{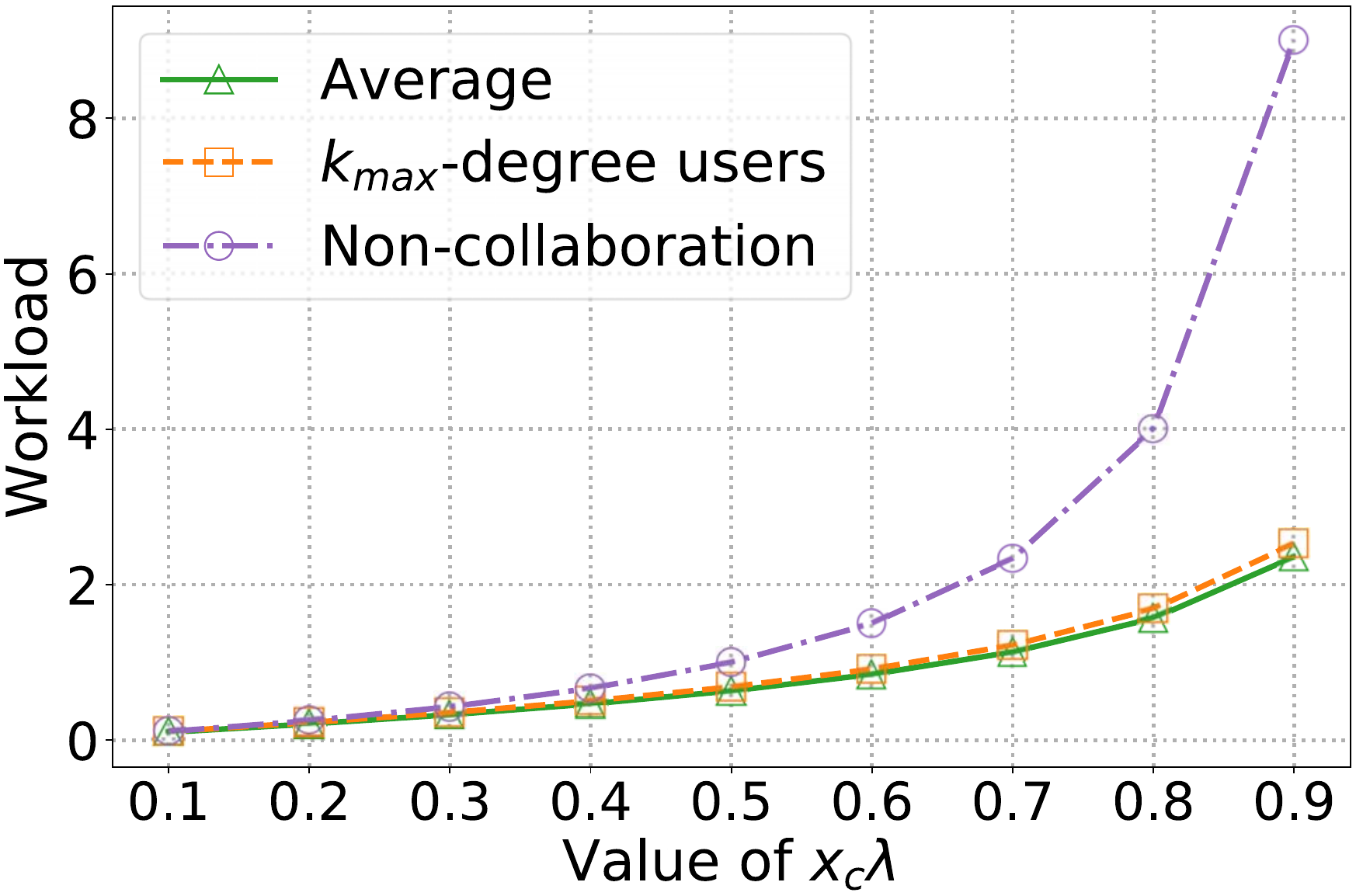}
    \caption{Workload comparison.}
    \label{Fig:max_MM1}
    \end{minipage}

    \begin{minipage}[t]{1.6in}
    \includegraphics[height=0.9in,width=1\columnwidth]{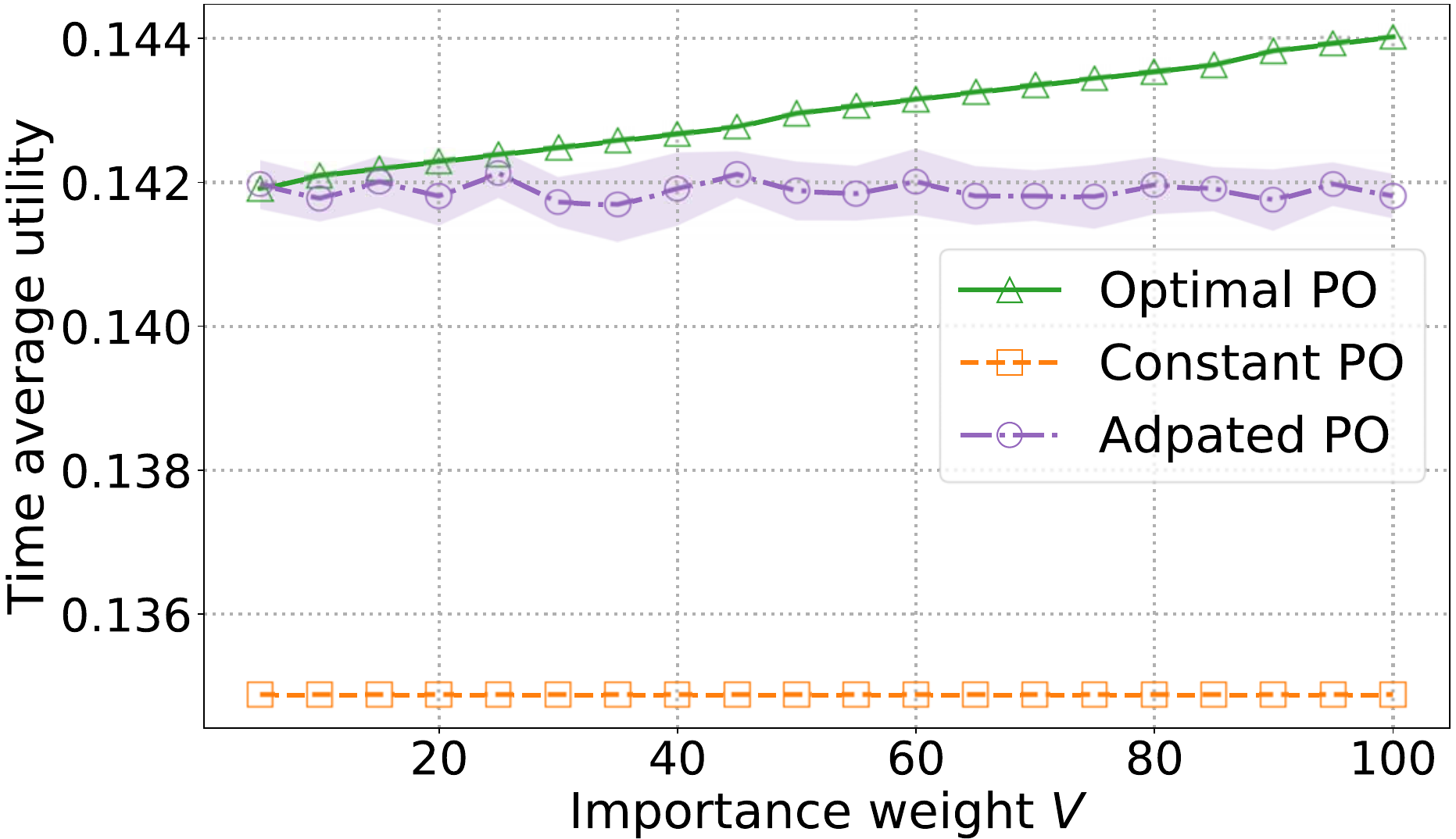}
    \caption{Utility vs. $V$.}
    \label{Fig:utilityV}
    \end{minipage}
    
    \begin{minipage}[t]{1.6in}
    \includegraphics[height=0.9in,width=1\columnwidth]{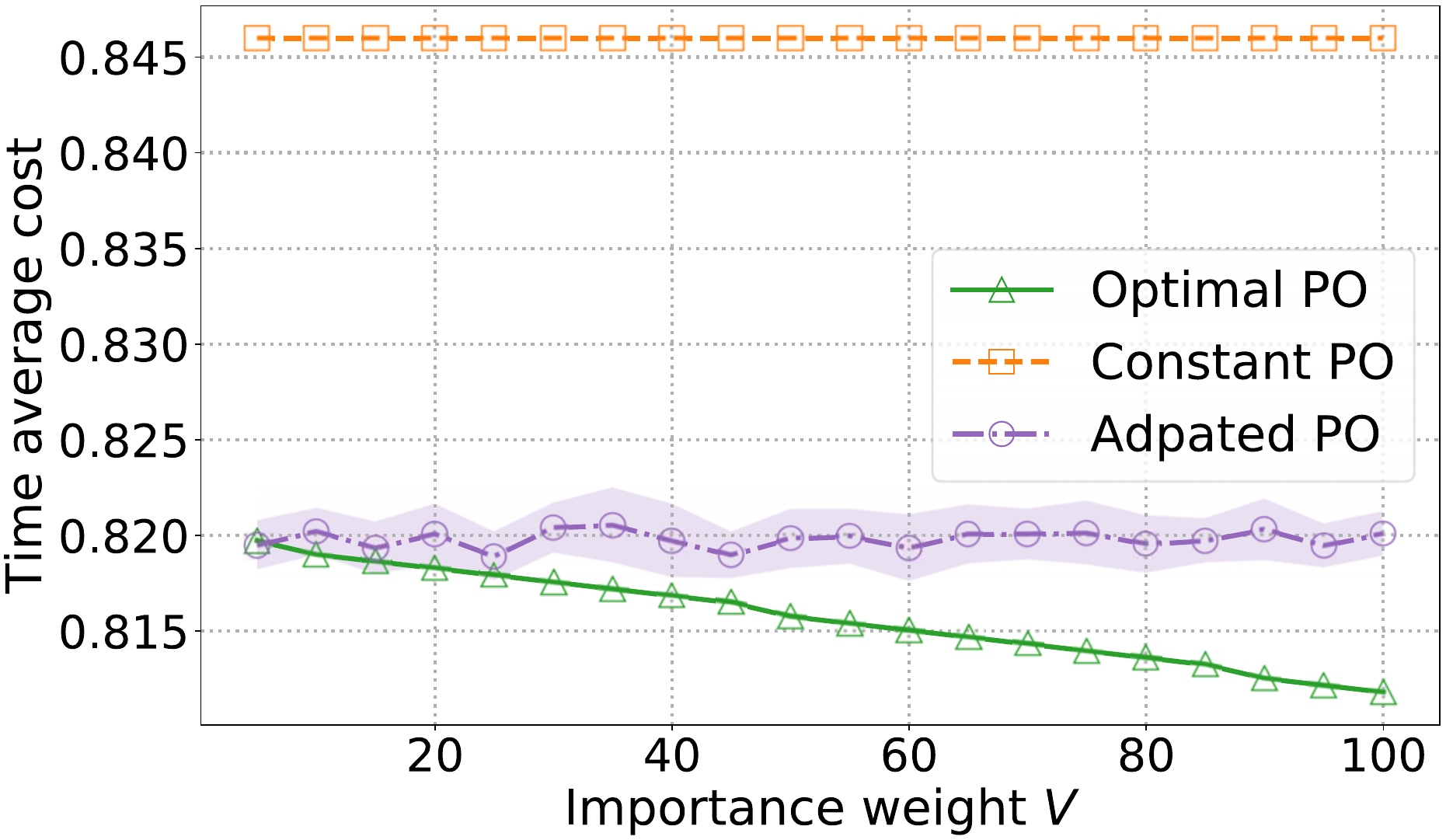}
    \caption{Cost vs. $V$.}
    \label{Fig:costV}
    \end{minipage}

    \begin{minipage}[t]{1.6in}
    \includegraphics[height=0.9in,width=1\columnwidth]{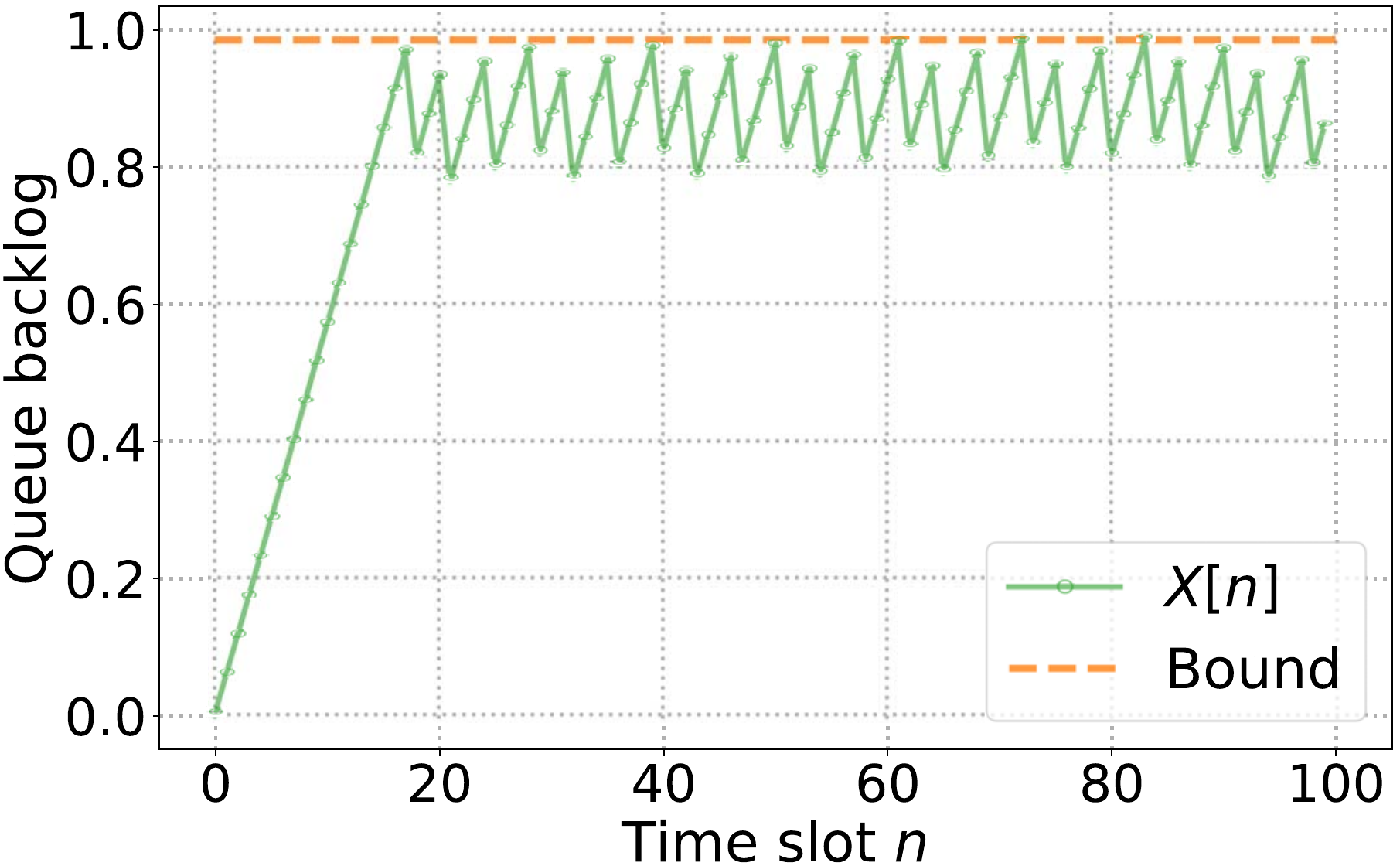}
    \caption{Queue backlog $X[n]$.}
    \label{Fig:queueXn}
    \end{minipage}

\end{tabular}
\vspace{-16pt}
\end{figure*}

\subsection{Mean Field D2D Collaboration}
\textbf{Stationary point}. We first demonstrate that the mean field model on graph is effective to characterize D2D collaboration by comparing the theoretical stationary point $\bm{s}^*$ obtained from the ODE system of Eq.~\eqref{Eq:ODE-s} and that from simulating the Po2 choices. In particular, we compute the theoretical stationary point $\bm{s}^*$ using \emph{scipy.integrate.odeint} in Python to solve the ODE system, since $\bm{s}(t)$ will converge to $\bm{s}^*$ when time $t$ is large enough. On the other hand, the simulated MEC system consists of $800$ users for static graph and $1000$ users for dynamic graph. The static graph is generated by the configuration model~\cite{pastor2015epidemic} with both self-loops and multiple edges between two users  being cut off to obtain an uncorrelated graph. Similarly, the configuration model is revoked when  the graph structure changes to produce dynamic graph. Varying the value $x_c \lambda$, namely the proportion of  tasks processed via D2D collaboration,  from $0.1$ to $0.9$ with an increment of $0.1$ each time, we run the simulated Po2 for eight times under each  $x_c \lambda$.  Figs.~\ref{Fig:MF_static}-\ref{Fig:MF_dynamic} exhibit the values of theoretical  $s^*_{k_{\min},1}$ and averaged simulated $s^*_{k_{\min},1}$ on static and dynamic graphs, respectively, which tell that theoretical results perfectly match with simulated results. Therefore, the mean field model is effective in analyzing D2D collaboration. 

Furthermore, we show each averaged simulated $s_{k,i}^*$ and theoretical $s_{k,i}^*$ when $x_c \lambda = 0.7$ in Table~\ref{Tab:compare}.  Results validate the accuracy of mean field model as the error between theoretical and simulated $s_{k,i}^*$ on static/dynamic graph is negligibly small. 

\begin{table}[t]
\centering
\caption{Theoretical and simulated stationary points.}\label{Tab:compare}
\begin{tabular}{|c||c|c|c|c|}
\hline
\textbf{$s_{k,i}^*$} & Theoretical &Static graph & Dynamic graph& Max error\\
\hline
\hline
$s_{6,1}^*$ &  0.66504 & 0.66706 & 0.67742 & 0.01238\\
\hline
$s_{7,1}^*$ &  0.68972 & 0.68924 & 0.696 & 0.00628\\
\hline
$s_{8,1}^*$ &  0.7123 & 0.71732 & 0.71786 & 0.00556\\
\hline
$s_{9,1}^*$ & 0.73295 & 0.73329 & 0.7325 & 0.00039\\
\hline
$s_{6,2}^*$ &  0.30585 & 0.31042 & 0.32056 & 0.01471\\
\hline
$s_{7,2}^*$ &  0.33239 & 0.34009 & 0.3492 & 0.01681\\
\hline
$s_{8,2}^*$ &  0.35814 & 0.35022 & 0.36905 & 0.00792\\
\hline
$s_{9,2}^*$ &  0.38302 & 0.36972 & 0.3852 & 0.0133 \\
\hline
\end{tabular}
\vspace{-15pt}
\end{table}

\textbf{User number influence}. We continue discussing how the simulated Po2 behaves over time for different number of users $N$. In particular, we display the evolution of $s_{k_{\min},1}(t)$ when $N=100,300,800$ for static graph and  $N=100,300,1000$ for dynamic graph in Figs.~\ref{Fig:MF_static_N} and \ref{Fig:MF_dynamic_N}, respectively. We can see that as $N$ increases largely, simulated results tend to  approach the theoretical stationary point with impaired variances.

\textbf{Power of collaboration}. Theorem~\ref{thm:heter_degree} reveals  that  largest-degree users have heaviest workloads. We now show that, even for those  users, their workloads are still effectively mitigated in  D2D collaboration, compared to the non-collaborative case which is a M/M/1 queue. Fig.~\ref{Fig:max_MM1} depicts the average workload and workloads for the largest-degree users as well as the non-collaborative case when value of $x_c\lambda$ varies. We can observe that the average and heaviest workloads are much smaller than the non-collaborative scenario, with the workload being mitigated by $73.8\%$ when $x_c\lambda=0.9$. Hence, the \emph{task delay is also significantly reduced} as a result of  the collaboration.
\vspace{-1pt}
\subsection{Lyapunov Optimization Based Offloading and Pricing}
\textbf{Critical points}. We first derive the critical points for task offloading. As for the delay constraint, we implement golden section search to numerically compute critical points $x_l^{*},x_u^{*}$ by using Eq.~\eqref{Eq:critical1}, and hence we obtain  $x_l^{*}=0.26586,x_u^{*}=0.72978$. Similarly, golden section search is leveraged to calculate critical points $x'_l,x'_u$ for the fairness constraint, and their values are $0.08505$ and $0.49953$, respectively. Overall, the feasible region for the offloading decision is $[0.49953, 0.72978]$, namely $x_l=0.49953, x_u=0.72978$.

\textbf{Utility and  cost}. Now we compare the performance of Lyapunov optimization (Optimal PO) with two baseline methods.
\begin{itemize}[leftmargin=*]
\item Constant PO: edge server always chooses the \underline{P}rice $p_u$ and users  react with the \underline{O}ffloading probability $x_l$.
\item Adapted PO: based on Eq.~\eqref{Eq:opt_price}, edge server chooses the  \underline{P}rice $\frac{\rho_c^m}{\mu^2} - \rho_t^m \frac{B}{r}$ or $p_u$ with probability $\frac{\overline{x}-x_l \lambda}{x_u \lambda-x_l \lambda}$ or $\frac{x_u \lambda-\overline{x}}{x_u \lambda-x_l \lambda}$, and users  react with the \underline{O}ffloading probability $x_u$ or $x_l$.
\end{itemize}
It can be verified that the overloaded constraint also holds for these baseline methods. Let the total time slot $T = 100$, and we vary the importance weight $V$ from $5$ to $100$  to obtain the corresponding  service utility and system cost.  Fig.~\ref{Fig:utilityV} displays the time average utility, which shows that the utility will increase over $V$  for Optimal PO as more emphasis is on the utility term. Besides, Optimal PO can achieve higher utility compared to Constant PO and Adapted PO. We then exhibit the time average system cost in Fig.~\ref{Fig:costV}, and we can see that Optimal PO leads to a lower cost than the baseline methods. With the increase of $V$, the edge server is more likely to set a lower price $\frac{\rho_c^m}{\mu^2} - \rho_t^m \frac{B}{r}$ from  Eq.~\eqref{Eq:opt_price}, that is why the system cost will decrease for Optimal PO. Therefore, Optimal PO can \emph{achieve high energy efficiency} in optimizing utility and cost.

\textbf{Queue backlog}. Lastly, we depict the queue backlog $X[n]$  in Fig.~\ref{Fig:queueXn} when the  importance weight is $V=20$. Results demonstrate that the queue backlog is finite under a constant bound derived in Theorem~\ref{thm:Lyapunov}. This also implies the overloaded constraint in utility maximization is satisfied.


\section{Related Work} \label{Sec:related}
\textbf{Collaborative MEC}. Emerging MEC offers new possibility for intelligent mobile applications~\cite{the2016satyanarayanan}. As single edge server has limited computing capacity, collaborative MEC is  effective to accommodate more computation~\cite{ning2018a}. Li \emph{et al.} propose an online learning aided collaborative offloading allowing edge servers to transmit tasks to each other based on a joint consideration of trust, delay and multi-hop transmission~\cite{li2019learning}. These works mainly focus on the collaboration among edge servers,  instead of, among mobile users. To explore how users can help each other, Pu \emph{et al.} study an incentive-aware task offloading among users via D2D links~\cite{pu2016D2D}, but offloading to edge server is not considered. He \emph{et al.} further incorporate D2D collaboration and task offloading to edge server for enhancing the computation capacity~\cite{he2019D2D}. However, existing works rarely investigate D2D collaboration and task offloading as a whole. Besides, they mainly concentrate on the centralized collaboration for finite, often a handful of,  users which makes them hard to be extended to large-population MEC systems. 

\textbf{Mean field model}. Mean field model is used  to characterize interactions among a large number of agents. Mitzenmacher uses the mean field model to analyze  the power of $d$ choices in randomized load balancing, where $d$ queues are  randomly sampled and a task will join the shortest queue~\cite{mitzenmacher2001the}.  Results show that even  $d=2$, the average sojourn time still decreases dramatically.  Later on, Gast investigates the power of two choices on finite-degree graphs, but only simulation results are provided~\cite{gast2015the}. Budhiraja \emph{et al.} show that the power of $d$ choices on graphs can still be analyzed via mean field model,  as long as each node has infinite degree~\cite{budhiraja2017super}.  Nevertheless, for  D2D collaboration,  the degree of a user (node) will not scale with the total  number of users due to short-range D2D communication, so that users actually have finitely many neighbors. Another strand of researches using mean field model on graphs focus on epidemic processes in networks~\cite{pastor2015epidemic}. These works usually assume an uncorrelated graph, and use an ODE system to represent the state evolution, whereas the convergence to mean field model is often not proved~\cite{farooq2019model}. Therefore, previous studies have not yet rigorously analyzed the mean field model on finite-degree or dynamic graphs.

\textbf{Service pricing}. Pricing scheme is important to a service provider when providing specific service for end users~\cite{zhang2014time}. Regarding MEC, Zhao \emph{et al.} propose a pricing scheme to charge mobile users when they offload computation via access points~\cite{zhao2020intelligent}. A Stackelberg game based  heterogeneous pricing is designed to make decisions for edge servers (leaders) and mobile users (followers) in~\cite{chen2020a}. However, these works only consider finitely many users, and their pricing schemes are too complicated as uniform pricing is more easily implementable.



\section{Conclusion}\label{Sec:conclusion}
In this paper, we develop a joint D2D collaboration and task offloading for a large-population MEC system. Specifically, to characterize the state evolution of D2D collaboration, we propose a mean field model to analyze the stochastic MEC system by a deterministic ODE system. On this basis,  we derive the existence and uniqueness of  the stationary point, and further demonstrate the global convergence of state evolution to this unique stationary point.  By incorporating D2D collaboration, we then  design a pricing scheme for task offloading following a Lyapunov optimization framework. In particular, the offloading process is modeled as a Stackelberg game,  where edge server is the leader to determine a reasonable price and users are followers to  make the offloading decision. Extensive evaluations validate the effectiveness of our mean field model and the superiority of Lyapunov optimization.

\section*{Acknowledgment}
This work is supported in part by the GRF 14201819 and CUHK:6905407.


\newpage
\begin{appendices}
\section{Details for Collaboration on Dynamic Graph}\label{App:dynamicG}
\textbf{Dynamic graph model}. The detailed derivation of D2D collaboration on dynamic graph is presented in this appendix. For dynamic graph, given  the expected degree $k, k \in \mathcal{K}$ of a user, its realized degree $k_1$ is distributed in the set $\mathcal{J}$ with probability conditioned on $k$, which is denoted as $\pi(k_1|k)$. One example for such conditional distribution is Poisson distribution for modeling random spatial networks~\cite{ lang2018analytic}. Also note that the realized degree set $\mathcal{J}$ could be different from the expected degree set $\mathcal{K}$. Due to time-varying D2D links, the realized degree $k_1$ is regenerated once the graph structure changes. For notional convenience, we use $p(k_2|k_1)$ to denote  the probability that a  user with realized degree $k_1$ has a neighbor with realized degree $k_2$, and $\pi(k'|k_2)$ to represent the probability when a user has realized degree $k_2$  while its expected degree is $k'$. Analogous to the static graph, we derive the evolution of each state $s_{k,i}$ from the perspective of a particular user $u$ with expected degree $k$. In the meantime, the transitions of  $s_{k,i}$ also include three instances which are the same as the static graph. A brief description is provided here for completeness: task generated and not offloaded stays at $u$; task sent from a neighbor of $u$; task processed by $u$.

\textbf{State evolution}. For the first instance where $u$ generates a  task and the task enters into $Q_u$, denote the corresponding probability as $p_1$. To obtain $p_1$, we need to consider two cases: 1-1) $u$ holds fewer tasks; 1-2) tie breaks. We will elaborate case 1-1) while directly providing the result for case 1-2) since they share a similar spirit. The probability of case 1-1) is $q_{k,i-1} \sum_{k_1 \in \mathcal{J}}\pi(k_1|k) \sum_{k_2 \in \mathcal{J}} p(k_2|k_1) \sum_{k' \in \mathcal{K}} \pi(k'|k_2) s_{k',i}$. The meaning of each term is explained: $\sum_{k_1 \in \mathcal{J}}\pi(k_1|k)$ is the probability $u$ has realized degree $k_1$, $\sum_{k_2 \in \mathcal{J}} p(k_2|k_1)$ means the probability that the polled neighbor has realized degree $k_2$, and $ \sum_{k' \in \mathcal{K}} \pi(k'|k_2)$ denotes the probability that the polled neighbor has expected degree $k'$. Then, we have:
 \begin{displaymath}
\begin{aligned}
&q_{k,i-1} \sum_{k_1 \in \mathcal{J}}\pi(k_1|k) \sum_{k_2 \in \mathcal{J}} p(k_2|k_1) \sum_{k' \in \mathcal{K}} \pi(k'|k_2) s_{k',i}\\
&= q_{k,i-1} \sum_{k_1 \in \mathcal{J}}\pi(k_1|k) \sum_{k_2 \in \mathcal{J}} \frac{k_2\sum_{k \in \mathcal{K}}\pi(k_2|k)p(k)}{\mathbb{E}[k_1]} \\
& \times \sum_{k' \in \mathcal{K}} \frac{\pi(k_2|k')p(k')}{\sum_{k' \in \mathcal{K}}\pi(k_2|k')p(k')}  s_{k',i}\\
& = q_{k,i-1} \sum_{k_1 \in \mathcal{J}}\pi(k_1|k) \sum_{k_2 \in \mathcal{J}} \frac{k_2  \sum_{k' \in \mathcal{K}} \pi(k_2|k')p(k')}{\mathbb{E}[k_1]}  s_{k',i}\\
& = q_{k,i-1}  \sum_{k_2 \in \mathcal{J}} \frac{k_2  \sum_{k' \in \mathcal{K}} \pi(k_2|k')p(k')}{\mathbb{E}[k_1]}  s_{k',i}\\
& =  q_{k,i-1}  \sum_{k' \in \mathcal{K}} \frac{p(k') \sum_{k_2 \in \mathcal{J}} k_2 \pi(k_2|k')}{\mathbb{E}[k_1]} s_{k',i}\\
& = q_{k,i-1}\sum_{k' \in \mathcal{K}} \frac{p(k') k'}{\overline{k}} s_{k',i}.
\end{aligned}
\end{displaymath}
Here, the first equality is due to the uncorrelated graph  $ p(k_2|k_1) =  \frac{k_2 p(k_2)}{\mathbb{E}[k_1]}=\frac{k_2\sum_{k \in \mathcal{K}}\pi(k_2|k)p(k)}{\mathbb{E}[k_1]}$. The last equality is because $\sum_{k_2 \in \mathcal{J}} k_2 \pi(k_2|k')=k'$ as $k'$ is the expected degree, and $\mathbb{E}[k_1] = \sum_{k \in \mathcal{K}}p(k) \sum_{k_1\in \mathcal{J}}\pi(k_1|k)k_1 = \sum_{k \in \mathcal{K}}p(k) k=\overline{k}$ is the expected degree over graph $\mathcal{G}(t)$. Analogously, we obtain the probability of case 1-2) as $q_{k,i-1}\frac{1}{2}\sum_{k' \in \mathcal{K}} \frac{p(k') k'}{\overline{k}} q_{k',i-1}$. Overall, $p_1 = q_{k,i-1}\sum_{k' \in \mathcal{K}} \frac{p(k') k'}{\overline{k}} s_{k',i} + q_{k,i-1}\frac{1}{2}\sum_{k' \in \mathcal{K}} \frac{p(k') k'}{\overline{k}} q_{k',i-1}$, which is the same as that of the first instance in static graph.

The probability of the second instance where $u$ receives a task sent from a neighbor is denoted as $p_2$, which also includes two cases: 2-1) $u$ has fewer tasks; 2-2) tie breaks. Based on the second instance in static graph and the first instance in dynamic graph, we attain the probability of case 2-1) below:
\begin{displaymath}
\scalebox{0.95}{$\begin{aligned}
& q_{k,i-1} \sum_{k_1 \in \mathcal{J}} \pi(k_1|k) k_1 \sum_{k_2 \in \mathcal{J}} p(k_2|k_1) \frac{1}{k_2} \sum_{k' \in \mathcal{K}}\pi(k'|k_2)s_{k',i}\\
& = q_{k,i-1} k \sum_{k_2 \in \mathcal{J}} \frac{k_2\sum_{k \in \mathcal{K}}\pi(k_2|k)p(k)}{\mathbb{E}[k_1] k_2}\sum_{k' \in \mathcal{K}} \frac{\pi(k_2|k')p(k') s_{k',i}}{\sum_{k' \in \mathcal{K}}\pi(k_2|k')p(k')} \\
& = q_{k,i-1} k  \sum_{k' \in \mathcal{K}} \frac{p(k')}{\overline{k}} s_{k',i}.
\end{aligned}$}
\end{displaymath} 
Following the same approach, we calculate the probability of case 2-2),  that is $q_{k,i-1}k  \frac{1}{2} \sum_{k' \in \mathcal{K}} \frac{p(k')}{\overline{k}} q_{k',i-1}$, and hence $p_2 = q_{k,i-1} k  \sum_{k' \in \mathcal{K}} \frac{p(k')}{\overline{k}} s_{k',i} + q_{k,i-1}k  \frac{1}{2} \sum_{k' \in \mathcal{K}} \frac{p(k')}{\overline{k}} q_{k',i-1}$, which is also the same as the second instance in static graph. In addition, the probability of the third instance is $q_{k,i}$. 

Integrating the three instances, we find out that the state evolution on dynamic graph  is exactly the ODE of  Eq.~\eqref{Eq:ODE-s}. 

\section{Proof of Existing Stationary point (Theorem~\ref{Thm:exist})}  \label{App:exist}

A stationary point is actually a fixed point such that:
\begin{displaymath}
\begin{aligned}
&x_c \lambda (s_{k,i-1}^*-s_{k,i}^*)\Bigl[\frac{1}{2}\sum_{k'\in\mathcal{K}}\frac{k'+k}{\overline{k}}p(k')(s_{k',i-1}^*+s_{k',i}^*)\Bigr]\\
&-\mu(s_{k,i}^*-s_{k,i+1}^*)=0.
\end{aligned}
\end{displaymath}
Let $z_{k,i} =  \frac{1}{2}\sum_{k'\in\mathcal{K}}\frac{k'+k}{\overline{k}}p(k')(s_{k',i-1}+s_{k',i})$, and then the RHS of  Eq.~\eqref{Eq:ODE-s} is rewritten as $x_c \lambda (s_{k,i-1}-s_{k,i}) z_{k,i} - \mu(s_{k,i}-s_{k,i+1})$. Accordingly, we define $G_{k,i}(\bm{s})$ which satisfies the following condition:
\begin{equation} \label{Eq:Gfunc}
 x_c \lambda G_{k,i}(\mathbf{s}) z_{k,i} +  \mu G_{k,i}(\mathbf{s}) - x_c \lambda s_{k,i-1}z_{k,i} -\mu s_{k,i+1}=0.
\end{equation}

\textbf{$G_{k,i}(\bm{s})$ is a function}. One should prove there is only one solution to Eq.~\eqref{Eq:Gfunc}, that is $G_{k,i}(\bm{s})$.  Construct a sequence of mapping $H_{k,i}(y_{k,i}) = x_c \lambda y_{k,i} z_{k,i} + \mu y_{k,i} - x_c \lambda s_{k,i-1}z_{k,i} - \mu s_{k,i+1}$. Hence, $H_{k,i}(s_{k,i-1}) =  \mu s_{k,i-1} - \mu s_{k,i+1} \ge 0 $ and $H_{k,i}(s_{k,i+1}) =  x_c \lambda s_{k,i+1}z_{k,i} - x_c \lambda s_{k,i-1}z_{k,i} \le 0$. Moreover,  $H_{k,i}(y_{k,i})$ is a monotonously increasing function, so that there is a unique solution to $H_{k,i}(y^*)=0$, that is $y^* = G_{k,i}(\bm{s})$.  Since $H_{k,i}(s_{k,i-1}) \ge 0$ and $H_{k,i}(s_{k,i+1}) \le 0$, we also have:
\begin{equation} \label{Eq:G_order}
s_{k,i-1} \ge G_{k,i}(\mathbf{s}) \ge s_{k,i+1}.
\end{equation}

\textbf{$\bm{G}(\bm{s})$ maps $\bm{s}$ to the same space}. Denote $\mathcal{S}$ as the space of $\bm{s}$. Any point $\bm{s} \in \mathcal{S}$ needs to satisfy $1 \ge s_{k,i} \ge s_{k,i+1} \ge 0$.  From Eq.~\eqref{Eq:G_order}, we have $G_{k,i}(\mathbf{s}) \ge s_{k,i+1} \ge 0$, and $G_{k,i}(\mathbf{s}) \le s_{k,i-1} \le 1$.  The remaining issue is to show  $G_{k,i}(\bm{s}) \ge G_{k,i+1}(\bm{s})$, which suffices to compare $ x_c \lambda G_{k,i}(\mathbf{s}) z_{k,i} +  \mu G_{k,i}(\mathbf{s})$ and $ x_c \lambda G_{k,i+1}(\mathbf{s}) z_{k,i} +  \mu G_{k,i+1}(\mathbf{s})$.
\begin{displaymath}
\begin{aligned}
 & x_c \lambda G_{k,i}(\mathbf{s}) z_{k,i} +  \mu G_{k,i}(\mathbf{s})-x_c \lambda G_{k,i+1}(\mathbf{s}) z_{k,i} -  \mu G_{k,i+1}(\mathbf{s})\\
 & =  x_c \lambda s_{k,i-1}z_{k,i} + \mu s_{k,i+1} -     x_c  \lambda s_{k,i}z_{k,i+1} - \mu s_{k,i+2}\\ 
& - x_c \lambda G_{k,i+1}(z_{k,i}-z_{k,i+1})\\
& \ge x_c \lambda s_{k,i-1}z_{k,i}  -    x_c \lambda s_{k,i}z_{k,i+1} - x_c \lambda G_{k,i+1}(z_{k,i}-z_{k,i+1})\\
& \ge x_c \lambda s_{k,i-1}(z_{k,i}-z_{k,i+1}) - x_c \lambda G_{k,i+1}(z_{k,i}-z_{k,i+1})\\
& \ge 0,
\end{aligned}
\end{displaymath}
where the first equality is from Eq.~\eqref{Eq:Gfunc}, the first inequality is due to $s_{k,i+1} \ge s_{k,i+2}$, the second inequality is because of $s_{k,i-1} \ge s_{k,i}$, and the last inequality is based on Eq.~\eqref{Eq:G_order}. As a result, we have $\bm{G}(\bm{s}) \in \mathcal{S}$.

\textbf{$G_{k,i}(\bm{s})$ is continuous in $\bm{s}$}.  For any two points $\bm{s}$ and $\bm{o}$, suppose $||\bm{s} - \bm{o}||_{\infty} \le \epsilon$. If $|G_{k,i} (\bm{s}) - G_{k,i}(\bm{o}) | \le C \epsilon$ with $C$ being a finite constant, then $G_{k,i}(\bm{s})$ is continuous. Use $\Delta = | x_c \lambda G_{k,i}(\bm{s})z_{k,i}(\bm{s}) +  \mu G_{k,i}(\bm{s}) -x_c \lambda G_{k,i}(\bm{o}) z_{k,i}(\bm{o}) -  \mu G_{k,i}(\bm{o})|$, and then:

\begin{displaymath}
\begin{aligned}
\Delta &=|(x_c \lambda z_{k,i}(\bm{s}) +\mu)(G_{k,i}(\bm{s}) - G_{k,i}(\bm{o}))\\
&+ x_c \lambda G_{k,i}(\bm{o})(z_{k,i}(\bm{s}) - z_{k,i}(\bm{o}))| \\
& \ge (x_c \lambda z_{k,i}(\bm{s}) +\mu)|G_{k,i}(\bm{s}) - G_{k,i}(\bm{o})|\\
&- x_c \lambda G_{k,i}(\bm{o})|z_{k,i}(\bm{s}) - z_{k,i}(\bm{o})|\\
& \ge \scalebox{0.93}{$(x_c \lambda z_{k,i}(\bm{s}) +\mu)|G_{k,i}(\bm{s})-G_{k,i}(\bm{o})|- x_c \lambda |z_{k,i}(\bm{s}) - z_{k,i}(\bm{o})|$}\\
& \ge \scalebox{0.98}{$ (x_c \lambda z_{k,i}(\bm{s}) +\mu)|G_{k,i}(\bm{s})-G_{k,i}(\bm{o})|- x_c \lambda \Bigl(1+\frac{k_{\max}}{\overline{k}}\Bigr) \epsilon$},
\end{aligned}
\end{displaymath}
where the second inequality is because $G_{k,i}(\bm{o}) \le 1$, and the last inequality is according to the definition of $z_{k,i}$. Furthermore, we replace $x_c \lambda G_{k,i}(\mathbf{s}) z_{k,i}(\mathbf{s}) +  \mu G_{k,i}(\mathbf{s})$ with $ x_c \lambda s_{k,i-1} z_{k,i}(\mathbf{s}) +  \mu s_{k,i+1}$, and  $x_c \lambda G_{k,i}(\mathbf{o}) z_{k,i}(\mathbf{o}) +  \mu G_{k,i}(\mathbf{o})$ with $x_c \lambda o_{k,i-1} z_{k,i}(\mathbf{o}) +  \mu o_{k,i+1}$. Similarly, we obtain:
\begin{displaymath}
\Delta  \le (x_c \lambda z_{k,i}(\mathbf{s}) +\mu)  \epsilon + x_c \lambda \Bigl(1+\frac{k_{\max}}{\overline{k}}\Bigr) \epsilon.
\end{displaymath}
By comparison, it yields $(x_c \lambda z_{k,i}(\mathbf{s}) +\mu)|G_{k,i} (\mathbf{s})-G_{k,i}(\mathbf{o})| \le (x_c \lambda z_{k,i}(\mathbf{s}) +\mu)  \epsilon + 2 x_c \lambda \left(1+\frac{k_{\max}}{\overline{k}}\right) \epsilon$. Dividing both sides by $(x_c \lambda z_{k,i}(\mathbf{s}) +\mu)$ and using  $x_c \lambda z_{k,i}(\mathbf{s}) +\mu \ge \mu$, we attain:
\begin{equation}
|G_{k,i} (\mathbf{s})-G_{k,i}(\mathbf{o})| \le \Bigl(1+ 2 x_c \lambda \frac{1}{\mu} + 2 x_c  \lambda \frac{k_{\max}}{\overline{k}\mu}\Bigr) \epsilon.
\end{equation}
Therefore,  if $\epsilon \rightarrow 0$, then $|G_{k,i} (\mathbf{s})-G_{k,i}(\mathbf{o})|  \rightarrow 0$, i.e., $G_{k,i}(\bm{s})$ is continuous in $\bm{s}$.

In general,  $G_{k,i}(\bm{s})$  is a continuous function which maps $\bm{s}$ to the same space. According to Brouwer fixed point theorem, there is a fixed point such that $G_{k,i}(\bm{s}^*) = s^*_{k,i}, \forall k,i$.  Because $G_{k,i}(\bm{s}^*)$ is the solution to Eq.~\eqref{Eq:Gfunc}, thus there exists a stationary point $\bm{s}^*$ for the mean field model of Eq.~\eqref{Eq:ODE-s}.

\section{Proof of Convergence and Uniqueness}
\subsection{Proof of Lemma~\ref{lem:dominance}} \label{App:dominance}
Considering that the solution $\bm{s}(t)$ is continuously dependent on the initial values, it is sufficient to prove the case where $s_{k,i}(0) > \hat{s}_{k,i}(0)$ and $s_{k,i}(t) \ge  \hat{s}_{k,i}(t), \forall k, i$. Suppose there is a critical time $t_1$ such that $s_{k,i}(t) > \hat{s}_{k,i}(t), \forall k, i$ when $t < t_1$ and $s_{k,i}(t_1) = \hat{s}_{k,i}(t_1)$ for some $k,i$. 

If $s_{k,i}(t_1) = \hat{s}_{k,i}(t_1), \forall k,i$. Obviously, $\bm{s}(t) = \bm{\hat{s}}(t), \forall t>t_1$, and the conclusion holds.

If  $\exists k',i'$ such that $s_{k',i'}(t_1) > \hat{s}_{k',i'}(t_1)$. Also there exist $k,i$ such that $s_{k,i}(t_1) = \hat{s}_{k,i}(t_1), i \ge 1$, and at least one of the following conditions holds: 1) $s_{k,i-1}(t_1) > \hat{s}_{k,i-1}(t_1)$, 2) $s_{k,i+1}(t_1) > \hat{s}_{k,i+1}(t_1)$. Still use $z_{k,i} =  \frac{1}{2}\sum_{k'\in\mathcal{K}}\frac{k'+k}{\overline{k}}p(k')(s_{k',i-1}+s_{k',i})$. From Eq.~\eqref{Eq:ODE-s}, we have:
\begin{equation}
\begin{aligned}
&\dot{s}_{k,i}(t_1)- \dot{\hat{s}}_{k,i}(t_1)\\
 &= \scalebox{0.96}{$x_c \lambda (s_{k,i-1}(t_1)-s_{k,i}(t_1))z_{k,i}(t_1)-\mu(s_{k,i}(t_1)-s_{k,i+1}(t_1))$}\\
 &- \scalebox{0.96}{$x_c \lambda (\hat{s}_{k,i-1}(t_1)-\hat{s}_{k,i}(t_1))\hat{z}_{k,i}(t_1)+\mu(\hat{s}_{k,i}(t_1)-\hat{s}_{k,i+1}(t_1))$}\\
 & = \scalebox{0.94}{$x_c \lambda (s_{k,i-1}(t_1)-s_{k,i}(t_1))z_{k,i}(t_1) + \mu (s_{k,i+1}(t_1) - \hat{s}_{k,i+1}(t_1))$}\\
 &- x_c \lambda (\hat{s}_{k,i-1}(t_1)-\hat{s}_{k,i}(t_1))\hat{z}_{k,i}(t_1)\\
 & =x_c \lambda (s_{k,i-1}(t_1)-s_{k,i}(t_1))(z_{k,i}(t_1)- \hat{z}_{k,i}(t_1))\\ 
&+ \scalebox{0.9}{$\mu (s_{k,i+1}(t_1) - \hat{s}_{k,i+1}(t_1)) + x_c \lambda \hat{z}_{k,i}(t_1) (s_{k,i-1}(t_1) - \hat{s}_{k,i-1}(t_1))$}\\
&>0.
\end{aligned}
\end{equation}
The last inequality is because at least 1) or 2) is true, and $z_{k,i}(t_1)>\hat{z}_{k,i}(t_1)$. Since both $s_{k,i}(t)$ and $\hat{s}_{k,i}(t)$ are continuous functions in time $t$, these must exist $t_0 < t_1$ which satisfies $s_{k,i}(t_0) > \hat{s}_{k,i}(t_0)$ and $\dot{s}_{k,i}(t)- \dot{\hat{s}}_{k,i}(t) > 0, \forall t \in (t_0,t_1)$. Recall from the definition of time derivation:
\begin{equation} 
\begin{aligned}
s_{k,i}(t_1) - \hat{s}_{k,i}(t_1) &= s_{k,i}(t_0)  -\hat{s}_{k,i}(t_0)\\
&+ \int_{t_0}^{t_1}(\dot{s}_{k,i}(t)- \dot{\hat{s}}_{k,i}(t)) dt > 0.
\end{aligned}
\end{equation}
which is contradictory to the assumption that $s_{k,i}(t_1) =\hat{s}_{k,i}(t_1)$. Therefore,  it is also true that $s_{k,i}(t_1) > \hat{s}_{k,i}(t_1), \forall k,i$.

Overall, when $\bm{s}(0) \succeq \bm{\hat{s}}(0)$, we have $\bm{s}(t) \succeq \bm{\hat{s}}(t), \forall t > 0$.

\subsection{Proof of Lemma~\ref{lem:convergence}} \label{App:convergence}

We will demonstrate that $\dot{\phi}(\mathbf{s}) < -\frac{1}{2}\phi(\mathbf{s})$. In particular, we mainly discuss the first case where $\bm{s}(0) \succeq \bm{s}^*, \forall \bm{s}^* \in \mathcal{S}^*$, while the second case is proved similarly. Besides, remove $ \min_{\bm{s}^* \in \mathcal{S}^*}$ and denote $\phi(\mathbf{s}) =\sum_{i \ge 0} \frac{|s_i -s_i^*|}{2^i}$ for brevity. Since $\bm{s}(0) \succeq \bm{s}^*$, then $\phi(\mathbf{s}) =\sum_{i \ge 0} \frac{\sum_{k \in \mathcal{K}} p(k)(s_{k,i}-s_{k,i}^*)}{2^i}=\sum_{i \ge 0} \frac{s_{i}-s_{i}^*}{2^i}$ according to Lemma~\ref{lem:dominance}. Note that  $s_{k,0}=1$ and $s_{(k),0}=\overline{k}$ for any $k \in \mathcal{K}$.  Using the fact that $\bm{s}^*$ is a stationary point, we have:
\begin{equation}
\scalebox{0.97}{$\begin{aligned}
\dot{\phi}(\mathbf{s}) &= \sum \nolimits_{i \ge 0} \frac{\dot{s}_i}{2^i}\\
&= \sum \nolimits_{i \ge 0} \frac{\frac{x_c \lambda}{\overline{k}}\left(s_{i-1}s_{(k),i-1}-s_is_{(k),i}\right)-\mu (s_i-s_{i+1})}{2^i}\\
& = \sum \nolimits_{i \ge 0} \frac{\frac{x_c \lambda}{\overline{k}}\left(s_{i-1}s_{(k),i-1}-s_is_{(k),i}\right)-\mu (s_i-s_{i+1})}{2^i} \\
&- \sum \nolimits_{i \ge 0} \frac{\frac{x_c \lambda}{\overline{k}}\left(s^*_{i-1}s_{(k),i-1}^{*}-s^*_i s_{(k),i}^{*}\right)-\mu(s^*_i-s_{i+1}^*)}{2^i}\\
& = - \sum \nolimits_{i \ge 1} \frac{x_c \lambda}{\overline{k}} \frac{s_is_{(k),i} - s^*_i s_{(k),i}^{*}}{2^{i+1}}  - \sum \nolimits_{i \ge 1} \frac{s_i - s^*_i}{2^{i+1}}\\
& \le - \frac{1}{2} \phi(\mathbf{s}),
\end{aligned}$}
\end{equation}
where the third equality is because $\bm{s}^*$ is a stationary point, and the inequality is based on Lemma~\ref{lem:dominance} and the definition of $s_i, s_{(k),i}$. For the case  where $\bm{s}^* \succeq \bm{s}(0), \forall \bm{s}^* \in \mathcal{S}^*$, we can also obtain $\dot{\phi}(\mathbf{s}) \le  - \frac{1}{2} \phi(\mathbf{s})$. Since $\phi(\mathbf{s}) = 0$ only at the stationary point, then $\bm{s}(t)$ will converge to the stationary points exponentially fast when $\bm{s}(0) \succeq \bm{s}^*, \forall \bm{s}^* \in \mathcal{S}^*$ or $\bm{s}^* \succeq \bm{s}(0), \forall \bm{s}^* \in \mathcal{S}^*$. Proof  completed.

\subsection{Proof of Theorem~\ref{thm:uniqueness}} \label{App:uniqueness}

We will draw this conclusion by first assuming there are multiple stationary points and then show a contradiction. 

Suppose the mean field model has at least two distinct stationary points $\bm{s}^*$ and $\bm{\hat{s}}^*$. Therefore, there exists a value $\epsilon >0$ such that the following distance  satisfies:
\begin{equation} \label{Eq:gap}
||\bm{s}^* - \bm{\hat{s}}^*|| = \sum_{i \ge 0} \frac{|\sum_{k \in \mathcal{K}} p(k)(s^*_{k,i}-\hat{s}_{k,i}^*)|}{2^i} \ge \epsilon.
\end{equation}
Let $\bm{s}(0) \succeq \bm{s}^*, \bm{\hat{s}}^*$, and hence $\bm{s}(t) \succeq \bm{s}^*, \bm{\hat{s}}^*$ by Lemma~\ref{lem:dominance}. Still use the distance definition in Eq.~\eqref{Eq:gap}, and it yields:
\begin{displaymath}
\epsilon \le ||\bm{s}^* - \bm{\hat{s}}^*|| \le  ||\bm{s}(t)-\bm{s}^*|| + ||\bm{s}(t)-\bm{\hat{s}}^*||.
\end{displaymath}
Denote $\phi_1(t) = ||\bm{s}(t)-\bm{s}^*||$ and $\phi_2(t) = ||\bm{s}(t)-\bm{\hat{s}}^*||$. Let $\phi(t) = \phi_1(t) + \phi_2(t)$. Using the same proving technique in Lemma~\ref{lem:convergence}, we attain that $\dot{\phi}_1(t) \le -\frac{1}{2} \phi_1(t)$ and  $\dot{\phi}_2(t) \le -\frac{1}{2} \phi_2(t)$, so that $\dot{\phi}(t) \le -\frac{1}{2} \phi(t) < 0$. As a result, these must exist a time $t_{\epsilon} >0$ such that $ \phi(t_{\epsilon}) < \epsilon$, which is  a contradiction. Hence, the mean field model has a unique stationary point.

\section{Proof  of Heterogeneous Degrees (Theorem~\ref{thm:heter_degree})}\label{App:heter_degree}
We first show the relation between $q_{k,i}^*$ and $q_{k',i}^*$. According to Eq.~\eqref{Eq:ODE-s}, the stationary point $\bm{s}^*$ satisfies:
\begin{displaymath}
x_c \lambda q^*_{k,i-1} z_{k,i}^* - \mu q_{k,i}^*=0,
\end{displaymath}
where $q^*_{k,i}$ is from Eq.~\eqref{Eq:Por-q} and $z_{k,i}^* = \frac{1}{2} \sum_{k_1 \in \mathcal{K}} \frac{k_1+k}{\overline{k}}p(k_1)(s^*_{k_1,i-1}+s^*_{k_1,i})$. Therefore:
 $$q_{k,i}^* = \frac{x_c \lambda}{\mu} z_{k,i}^* q^*_{k,i-1}.$$
 For any $k>k'$, suppose that $q^*_{k,0} \ge q^*_{k',0}$. Since $ z_{k,i}^* > z_{k',i}^*, \forall i \ge 0$,  we have $q^*_{k,1} > q^*_{k',1}$. By induction, $q^*_{k,i} > q^*_{k',i}, \forall i>0$. On the other side, $s_{k,0}^* = \sum_{i \ge 0} q_{k,i}^* = 1$ and $s_{k',0}^* = \sum_{i \ge 0} q_{k',i}^* = 1$, which is a contradiction. Hence, $q^*_{k,0} < q^*_{k',0}$. As $s_{k,0}^* = s_{k',0}^*=1$, there must exist an index $i_{k,k'}$ such that $q^*_{k,i} \le q^*_{k',i}, \forall i < i_{k,k'}$ and  $q^*_{k,i} > q^*_{k',i}, \forall i \ge  i_{k,k'}$, i.e., once $q^*_{k,i} > q^*_{k',i}$ then $q^*_{k,i'} > q^*_{k',i'}, \forall i' >i$.

Remember that  $s^*_{k,i} = \sum_{i' \ge i} q^*_{k,i'}$, and then  $s^*_{k,i} > s^*_{k',i}, \forall i \ge i_{k,k'}$.  As for $i < i_{k,k'}$, assume there exists an index $i^*$ which makes $s^*_{k,i^*} < s^*_{k',i^*}$. Because $s^*_{k,0} = \sum_{i=0}^{i^*-1} q^*_{k,i} + s^*_{k,i^*}$, $s^*_{k',0} = \sum_{i=0}^{i^*-1} q^*_{k',i} + s^*_{k',i^*}$, and $q^*_{k,i} < q^*_{k',i}, \forall i < i^*$, we have $s^*_{k,0} < s^*_{k',0}$, which is in contrast to the fact that $s^*_{k,0}=s^*_{k',0}=1$. Consequently, no such index $i^*$ exists, or $s^*_{k,i} \ge s^*_{k',i}, \forall i \ge 0$. Proof completed.

\section{Proof of Convergence to Mean Field Model}
\subsection{Proof of Lemma~\ref{lem:drift}} \label{App:drift}

In this proof, $||\cdot||_{\infty}$-norm is used  to measure the distance. Assume that $||\bm{s}-\bm{\hat{s}}||_{\infty} = d$, and we will show that $||\bm{F}(\bm{s}) - \bm{F}(\bm{\hat{s}})||_{\infty} \le Cd$. Still denote $z_{k,i} = \frac{1}{2} \sum_{k' \in \mathcal{K}} \frac{k'+k}{\overline{k}}p(k')(s_{k',i-1}+s_{k',i})$. Using Eq.~\eqref{Eq:ODE-s}, we have:
\begin{displaymath}
\begin{aligned}
& |F_{k,i}(\bm{s}) - F_{k,i}(\bm{\hat{s}})|\\
& = |x_c\lambda q_{k,i-1}z_{k,i} - \mu q_{k,i} - x_c \lambda \hat{q}_{k,i-1}\hat{z}_{k,i}+\mu \hat{q}_{k,i}|\\
& = |x_c \lambda z_{k,i}(q_{k,i-1} - \hat{q}_{k,i-1}) + x_c \lambda \hat{q}_{k,i-1}(z_{k,i}-\hat{z}_{k,i})\\
& - \mu(q_{k,i}-\hat{q}_{k,i})|\\
& \le x_c \lambda z_{k,i} |q_{k,i-1} - \hat{q}_{k,i-1}| + x_c \lambda \hat{q}_{k,i-1} |z_{k,i}-\hat{z}_{k,i}|\\
&+\mu|q_{k,i}-\hat{q}_{k,i}|\\
& \le x_c \lambda \Bigl(1+ \frac{k_{\max}}{\overline{k}}\Bigr) |q_{k,i-1} - \hat{q}_{k,i-1}| + x_c \lambda |z_{k,i}-\hat{z}_{k,i}|\\
&+\mu|q_{k,i}-\hat{q}_{k,i}|\\
& \le 2 x_c \lambda \Bigl(1+ \frac{k_{\max}}{\overline{k}}\Bigr)  d + x_c \lambda \Bigl(1+ \frac{k_{\max}}{\overline{k}}\Bigr)d +2\mu d\\
& = \Bigl[3x_c \lambda \Bigl(1+ \frac{k_{\max}}{\overline{k}}\Bigr) +2\mu \Bigr] d.
\end{aligned}
\end{displaymath}
The second inequality is from the definition of $z_{k,i}$ and the fact $s_{k,i}, q_{k,i} \le 1$. The third inequality is because $|q_{k,i-1}-\hat{q}_{k,i-1}| \le |s_{k,i-1}-\hat{s}_{k,i-1}| + |s_{k,i}-\hat{s}_{k,i}| \le d+d =2d$. Let $C = 3x_c \lambda \big(1+ \frac{k_{\max}}{\overline{k}}\big) +2 \mu$, we complete the proof.

\subsection{Proof of Theorem~\ref{thm:efficacy}} \label{App:efficacy}

The result is based on the  Kurtz's theorem.

\textbf{Density dependent process}. Since the graph $\mathcal{G}$ is connected and uncorrelated, it can be validated that $\bm{s}^{(N)}(t)$ is a density dependent jump Markov process in the state space  $\mathcal{S}$.

\textbf{Bounded transition rate}. Denote  the number of users with degree $k$ as $N_k$. At state $\bm{s}$, the transitions of the MEC system are given by $\mathcal{L}=\{\pm \bm{e}_{k,i}, k \in \mathcal{K}, i \ge 0 \}$, where $\bm{e}_{k,i}$ is a vector with  element corresponding to degree $k$ and task number $i$ equal to $\frac{1}{N_k}$, while others being $0$.  The transition rate of $+\bm{e}_{k,i}$ is calculated as $N_k x_c \lambda (s_{k,i-1}-s_{k,i})\left[\frac{1}{2}\sum_{k'\in\mathcal{K}}\frac{k'+k}{\overline{k}}p(k')(s_{k',i-1}+s_{k',i})\right]$ and that of $-\bm{e}_{k,i}$ is $ N_k \mu (s_{k,i} - s_{k,i+1})$. Therefore, the rate at which jumps occur is bounded above by $x_c \lambda \frac{\overline{k}+k_{\max}}{\overline{k}}+\mu$ everywhere.

\textbf{Lipschitz drift function}. The limiting mean field model is a deterministic process, described by the ODE system of Eq.~\eqref{Eq:ODE-s}.  Lemma~\ref{lem:drift} states that Eq.~\eqref{Eq:ODE-s} satisfies the Lipschitz condition.

Based on the Kurtz's theorem, Eq.~\eqref{Eq:Ninfty} holds almost surely.

\section{Proof of Stationary Point Discussion}
\subsection{Proof of Corollary~\ref{cor:bound}} \label{App:bound}

We will only prove the upper bound in Eq.~\eqref{Eq:upper}, while the lower bound is acquired following a similar approach.

From Theorem~\ref{thm:heter_degree}, we obtain $s^*_{k,i} \le s^*_{k_{\max},i}, \forall k \in \mathcal{K}$. Besides, the stationary point $\bm{s}^*$ satisfies:
\begin{displaymath}
\begin{aligned}
&\dot{s}^*_{k,i}= - \mu (s^*_{k,i} - s^*_{k,i+1})\\
&+ x_c \lambda (s_{k,i-1}^*-s_{k,i}^*)\Bigl[\frac{1}{2}\sum_{k'\in\mathcal{K}}\frac{k'+k}{\overline{k}}p(k')(s_{k',i-1}^*+s_{k',i}^*)\Bigr].
\end{aligned}
\end{displaymath}
In particular, let $k = k_{\max}$. Also, we denote $z^*_{k_{\max},i}=\frac{1}{2}\sum_{k'\in\mathcal{K}}\frac{k'+k_{\max}}{\overline{k}}p(k')(s_{k',i-1}^*+s_{k',i}^*)$ and $z^*_{(k_{\max}),i} = \frac{1}{2}\sum_{k'\in\mathcal{K}}\frac{k'+k_{\max}}{\overline{k}}p(k')(s_{k_{\max},i-1}^*+s_{k_{\max},i}^*)$. Therefore:
\begin{displaymath}
\begin{aligned}
&\dot{s}^*_{k_{\max},i} =  - \mu (s^*_{k_{\max},i} - s^*_{k_{\max},i+1} )\\
&+  x_c \lambda (s_{k_{\max},i-1}^*-s_{k_{\max},i}^*) z^*_{k_{\max},i}\\
&\le \scalebox{0.97}{$- \mu (s^*_{k_{\max},i} - s^*_{k_{\max},i+1} ) +  x_c \lambda (s_{k_{\max},i-1}^*-s_{k_{\max},i}^*)z^*_{(k_{\max}),i}$}\\
&=-\mu \left(s_{k_{\max},i}^*-s_{k_{\max},i+1}^* \right)\\
&+ x_c\lambda \frac{1+\delta_1}{2} \Bigl[\left(s_{k_{\max},i-1}^* \right)^2-\left(s_{k_{\max},i}^* \right)^2 \Bigr].
\end{aligned}
\end{displaymath}
The upper bound will converge to $ \left (\frac{1+\delta_1}{2} \frac{x_c \lambda}{\mu} \right)^{2^i -1}$, so that $s_{k,i}^*$ is bounded by  Eq.~\eqref{Eq:upper}. Proof completed.

\subsection{Proof of Corollary~\ref{cor:busy}} \label{App:busy}

Since the system is stable at the stationary point, task completion rate should be equal to task arrival rate. Formally, task generation rate  is $\lambda$, and a task is processed via D2D collaboration with  probability $x_c$. Besides, task completion  rate is $\sum_{k \in \mathcal{K}} p(k)s^*_{k,i} \mu = s^*_1 \mu$. As a result, we have $x_c \lambda = s^*_1 \mu$, which implies  
$s^*_1  = \frac{x_c \lambda}{\mu}$.

\subsection{Proof of Corollary~\ref{cor:relation}} \label{App:relation}

We show the result by a simple induction. Because $s_{1}^* = \frac{x_c \lambda}{\mu}$ according to Corollary~\ref{cor:busy}, and $s_0^* =1, s^{*}_{(k),0} = \overline{k}$. The equality holds when $i = 1$. Suppose the equality holds  for $i =i', i'>1$. We demonstrate that it is also true when  $i = i'+1$. From Eq.~\eqref{Eq:inv_sk}, since $s_{i'}^* =  \frac{x_c \lambda}{\overline{k} \mu} s^*_{i'-1} s^{*}_{(k),i'-1}$, it is intuitive that $s_{i'+1}^* =  \frac{x_c \lambda}{\overline{k} \mu} s^*_{i'} s^{*}_{(k),i'}$. Proof completed.

\section{Proof of Lyapunov Optimization}
\subsection{Proof of Lemma~\ref{lem:drift_bound}} \label{App:drift_bound}

The proof simply uses the fact that $[\max (a,0)]^2 \le a^2$. As a result, Lyapunov drift satisfies:
\begin{displaymath}\scalebox{0.92}{$
\begin{aligned}
& \frac{1}{2}X^2[n+1]  - \frac{1}{2}X^2[n] \le  \frac{1}{2}(X[n]+x(p[n])\lambda -\overline{x})^2 - \frac{1}{2}X^2[n] \\
&=  \frac{1}{2} X^2[n] + X[n](x(p[n])\lambda-\overline{x}) + \frac{1}{2} (x(p[n])\lambda -\overline{x})^2 - \frac{1}{2}X^2[n]\\
& =X[n](x(p[n])\lambda-\overline{x}) + \frac{1}{2} (x(p[n])\lambda -\overline{x})^2.
\end{aligned}$}
\end{displaymath}
Because $x(p[n]) \in [0,1]$, we have $x(p[n])\lambda -\overline{x} \in [-\overline{x}, \lambda-\overline{x}]$, so that $(x(p[n])\lambda -\overline{x})^2 \le \max \left((\lambda-\overline{x})^2, \overline{x}^2 \right)$. Combining with the utility in  Eq.~\eqref{Eq:utility}, we complete the proof.

\subsection{Proof of Theorem~\ref{thm:Lyapunov}}\label{App:Lyapunov}
Prior to proving Theorem~\ref{thm:Lyapunov}, we need to introduce a lemma from~\cite{neely2010stochastic}. 
\begin{lem} \label{lem:aid}
 Considering D2D collaboration among users, there exists  a stationary, randomized scheduling policy that makes decision $p^*[n] \in (0,p_u]$ in every time slot, and yields steady-state values:
 \begin{equation} \label{Eq:aid}
 \begin{aligned}
& \mathbb{E}[u[n]] = u^*,\\
& \mathbb{E}[x^*[n] \lambda] \le \overline{x},
 \end{aligned}
 \end{equation}
 where $u^*$ is the optimal time average service utility.
 \end{lem}
 Lemma~\ref{lem:aid} points out that there exists a randomized policy which can attain the optimal service utility. In the following, we present the proof of conclusions a) and b) sequentially.

 a) For the initial queue backlog  $X[0]$, the inequality Eq.~\eqref{Eq:Xn_bound} holds as $X[0] =0$. Next, we demonstrate that when  Eq.~\eqref{Eq:Xn_bound} holds in time slot $n$, it will be also satisfied in time slot $n+1$. Regarding the condition $X[n] \le  \frac{V x_u(\frac{\rho_c^m}{\mu^2} - \rho_t^m \frac{B}{r})-Vx_l p_u}{ x_u-x_l} -V\frac{\rho_c^s}{\gamma}  + x_u \lambda - \overline{x}$, it can be classified into two cases: 1) $X[n] \le  \frac{V x_u(\frac{\rho_c^m}{\mu^2} - \rho_t^m \frac{B}{r})-Vx_l p_u}{ x_u-x_l} -V\frac{\rho_c^s}{\gamma}$; 2) $\frac{V x_u(\frac{\rho_c^m}{\mu^2} - \rho_t^m \frac{B}{r})-Vx_l p_u}{ x_u-x_l} -V\frac{\rho_c^s}{\gamma} \le X[n] \le  \frac{V x_u(\frac{\rho_c^m}{\mu^2} - \rho_t^m \frac{B}{r})-Vx_l p_u}{ x_u-x_l} -V\frac{\rho_c^s}{\gamma} + x_u \lambda - \overline{x}$. For the first case $X[n] \le  \frac{V x_u(\frac{\rho_c^m}{\mu^2} - \rho_t^m \frac{B}{r})-Vx_l p_u}{ x_u-x_l} -V\frac{\rho_c^s}{\gamma}$, it is natural that $X[n+1] \le \frac{V x_u(\frac{\rho_c^m}{\mu^2} - \rho_t^m \frac{B}{r})-Vx_l p_u}{ x_u-x_l} -V\frac{\rho_c^s}{\gamma} + x_u \lambda -\overline{x}$ according to the queue dynamics in Eq.~\eqref{Eq:queue} since $x[n] \le x_u$. As for the second case, we combine the optimal price and the offloading decision to derive the result. Based on Eq.~\eqref{Eq:opt_price}, the price will be $p_u$ so that the offloading decision $x[n] = x_l$. Because $x_l <x_u$, and then $X[n+1] \le X[n] \le \frac{V x_u(\frac{\rho_c^m}{\mu^2} - \rho_t^m \frac{B}{r})-Vx_l p_u}{ x_u-x_l} -V\frac{\rho_c^s}{\gamma}  + x_u \lambda - \overline{x}$.
 
 b) The upper bound of drift-minus-utility is minimized in each time slot by  choosing the price based on Eq.~\eqref{Eq:opt_price}.  In line with Lemma~\ref{lem:drift_bound} and Lemma~\ref{lem:aid}, we have:
 \begin{equation} \label{Eq:Lya_Xn}
 \begin{aligned}
 & \Delta(X[n]) - V\mathbb{E}[u[n]|X[n]]\\
 & \le \mathbb{E}[X[n](x(p[n]) \lambda - \overline{x}|X[n]]  - V\mathbb{E}[u[n]] +D\\
 & \le  \mathbb{E}[X[n](x(p^*[n]) \lambda - \overline{x}|X[n]]  - V\mathbb{E}[u^*[n]] +D\\
 & \le D - Vu^*.
 \end{aligned}
 \end{equation}
 Take expectation on both sides over $X[n]$, and it yields:
 \begin{equation}
 \mathbb{E} \Bigl[ \frac{1}{2}X^2[n+1] \Bigr] -  \mathbb{E} \Bigl[ \frac{1}{2}X^2[n] \Bigr] - V\mathbb{E}[u[n]] \le D-Vu^*.
 \end{equation}
 Sum the equation from $n=0$ to $T-1$:
  \begin{equation}
 \mathbb{E} \Bigl[ \frac{1}{2}X^2[T] \Bigr] -  \mathbb{E} \Bigl[ \frac{1}{2}X^2[0] \Bigr] - V \sum_{n=0}^{T-1} \mathbb{E}[u[n]] \le TD-TVu^*.
 \end{equation}
 Divide by $VT$, and arrange the terms:
 \begin{equation}
 \begin{aligned}
 &\frac{1}{T}  \sum_{n=0}^{T-1} \mathbb{E}[u[n]]\\
  &\ge u^* -\frac{D}{V} + \mathbb{E} \Bigl[ \frac{1}{2VT}X^2[n+1] \Bigr] -  \mathbb{E} \Bigl[ \frac{1}{2VT}X^2[0] \Bigr]\\
 & \ge  u^* -\frac{D}{V}-  \mathbb{E} \Bigl[ \frac{1}{2VT}X^2[0] \Bigr].
\end{aligned}
 \end{equation}
 Since $X^2[0] = 0$, and let $T \rightarrow \infty$, we obtain:
 $$\lim_{T \rightarrow \infty} \frac{1}{T} \sum_{n=0}^{T-1} \mathbb{E}[u[n]]  \ge u^* - \frac{D}{V}.$$
 Part b) is proved.

\end{appendices}

\end{document}